\documentclass[prd,showpacs,twocolumn,superscriptaddress,floatfix,10pt]{revtex4-2}
\usepackage{setspace} 
\usepackage[utf8]{inputenc}
\usepackage{float}
\usepackage{amsmath,amssymb,amsfonts,bm}
\usepackage{graphicx}
\usepackage{multirow}
\usepackage[normalem]{ulem}
\usepackage[usenames,dvipsnames]{color}
\allowdisplaybreaks
\usepackage[
colorlinks=true,
linkcolor=blue,
breaklinks=true,
urlcolor=blue,
citecolor=blue]{hyperref}

\usepackage{orcidlink}
 
\definecolor{green}{rgb}{0,0.6,0}

\newcommand{\lhc}{{\rm lhc}}
\newcommand{\rhc}{{\rm rhc}}

\newcommand{\be}{\begin{equation}} 
\newcommand{\ee}{\end{equation}}
\newcommand{\bea}{\begin{eqnarray}} 
\newcommand{\eea}{\end{eqnarray}}
\newcommand{\beas}{\begin{eqnarray*}} 
\newcommand{\eeas}{\end{eqnarray*}}

\newcommand{\rub}{\affiliation{Institut f\"ur Theoretische Physik II, Ruhr-Universit\"at Bochum, D-44780 Bochum, Germany }}

\begin{document}
\title{Solving the left-hand cut problem in  lattice QCD:\\
 $T_{cc}(3875)^+$ from finite volume energy levels }

\begin{abstract}
We discuss a novel effective-field-theory-based approach for extracting two-body scattering information from finite volume energies, serving as an alternative to Lüscher's method. By explicitly incorporating one-pion exchange, we overcome the challenging left-hand cut problem in L\"uscher’s method  and can handle finite volume energy levels both below and above the left-hand cut. Applied to the lattice data for $DD^*$ scattering at a pion mass of 280 MeV,  as an illustrative example, our results reveal the significant impact of the one-pion exchange on P-wave and S-wave phase shifts. The pole position of the $T_{cc}(3875)^+$ state, extracted from  the
finite-volume energy levels at this pion mass while taking into account  left-hand cut effects, range corrections and partial-wave mixing, is consistent with a near-threshold resonance. 
This study demonstrates, for the first time,  that  two-body scattering information can be reliably extracted from lattice spectra including the left-hand cut. 
\end{abstract}

\author{Lu Meng \orcidlink{0000-0001-9791-7138}}
\rub

\author{Vadim Baru\orcidlink{0000-0001-6472-1008}}
\rub

\author{Evgeny Epelbaum\orcidlink{0000-0002-7613-0210}}
\rub

\author{Arseniy~A.~Filin\orcidlink{0000-0002-7603-451X}}
\rub

\author{Ashot~M.~Gasparyan\orcidlink{0000-0001-8709-4537}}
\rub

\maketitle

\section{Introduction}
Over the last two decades, numerous exotic hadronic states have been discovered in the heavy quark sector, challenging conventional quark models. Quantum chromodynamics (QCD), with its color confinement,  is compatible with a wide range of color-neutral hadrons, such as multiquarks, 
hadronic molecules, hybrids, glueballs etc., see \cite{Esposito:2016noz,Lebed:2016hpi,Guo:2017jvc,Yamaguchi:2019vea,Brambilla:2019esw,Guo:2019twa,Chen:2022asf,Meng:2022ozq} for the review articles. 
Yet, the specific configurations that are realized in nature remain enigmatic. Consequently, experimental searches for exotic hadrons and the analysis of data in a manner consistent with unitarity and analyticity, allowing for the appropriate extraction of pole positions, are fundamental for enhancing our understanding of the strong interaction  
in the Standard Model.
 Additionally, the pertinent information can be gained from lattice simulations -- a first principle approach to solve QCD in a non-perturbative regime, see \cite{Briceno:2017max,Aoki:2020bew,Mai:2021lwb,Bicudo:2022cqi,Bulava:2022ovd,Prelovsek:2023sta} for recent reviews.

Recently, LHCb observed the first manifestly exotic doubly-charmed narrow resonance $T_{cc}(3875)^+$,  whose minimal quark content is $cc\bar u \bar d$~\cite{LHCb:2021vvq,LHCb:2021auc}. 
With its mass  being just a few hundreds keV below the $D^{*+}D^0$ threshold and the width  almost completely dominated
 by the only available strong decay to $DD\pi$, this state has been extensively analyzed using low-energy effective field theories (EFT)   \cite{Albaladejo:2021vln,Meng:2021jnw,Du:2021zzh,Braaten:2022elw,Wang:2022jop,Dai:2023mxm} and phenomenological models, see, e.g.,  
 \cite{Chen:2022asf} and references therein.

The $T_{cc}$ has also been recently investigated in lattice QCD \cite{Padmanath:2022cvl,Chen:2022vpo,Lyu:2023xro}. 
In the first two studies, the L\"uscher method was employed to determine the $DD^*$ phase shifts   (step 1)
at pion masses of 280 and 350 MeV, respectively.
The extracted infinite-volume phase shifts were then 
parameterized using the effective-range expansion (ERE)   (step 2),  leading to the determination of low-energy parameters for $DD^*$ scattering, namely the scattering length and effective range. Furthermore,  the pole position determined in Ref.~\cite{Padmanath:2022cvl} is consistent with the $T_{cc}$ being a virtual state, indicative of its molecular nature \cite{Matuschek:2020gqe}.
However, the analyses of Refs.~\cite{Padmanath:2022cvl,Chen:2022vpo} were questioned in a recent study \cite{Du:2023hlu}, which  highlighted the important role of the one-pion exchange (OPE), which brings a new scale into the problem from a nearby left-hand cut (lhc). 
The presence of the lhc restricts the ERE, commonly used for analyzing infinite volume phase shifts at   step 2, to a very narrow energy range, rendering it unsuitable for accurate pole extractions~\cite{Du:2023hlu}.
Moreover, the validity of the L\"uscher formula~\cite{Luscher:1986pf,Luscher:1990ux,Kim:2005gf,Briceno:2014oea}, 
which is the cornerstone for extracting infinite volume amplitudes from finite volume (FV) energy levels  at step 1, becomes questionable in the presence of  a nearby lhc~\cite{Raposo:2023oru,Dawid:2023jrj,Green:2021qol}.

\begin{figure}[tp]
\begin{center}
\includegraphics[width=0.48\textwidth]{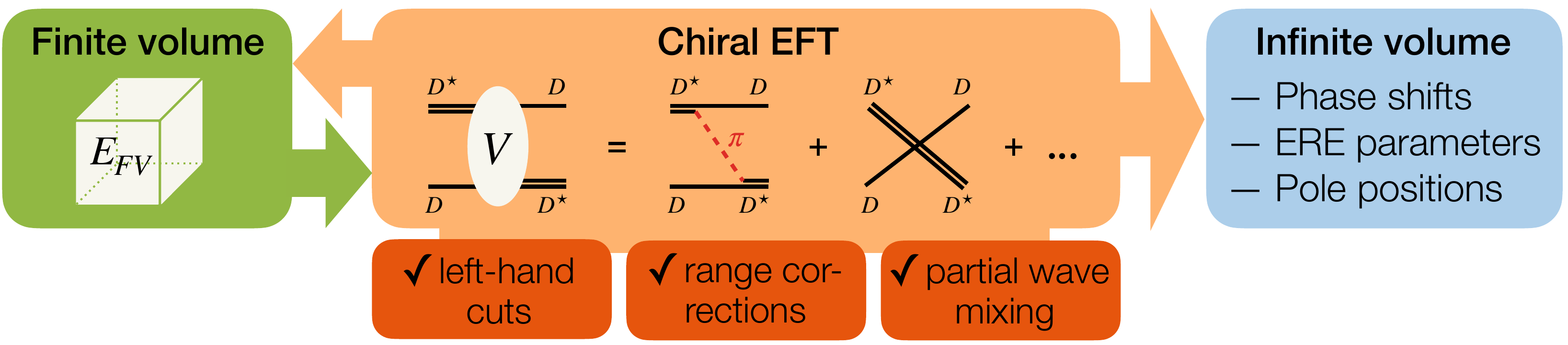}
\caption{Schematic illustration of the approach employed in this study.  
$V$  denotes the effective potential  in chiral EFT, involving the OPE and contact interactions;
$E_{FV}$ stands for the finite volume energy levels in lattice simulations, used here as input. 
}
\label{fig:our_method}
\end{center}
\end{figure}

In this study, we  resolve the challenging lhc problem inherent in Lüscher's method, which is fundamental for extracting two-body scattering information from finite volume energy spectra. 
This achievement is made possible by employing a chiral EFT-based approach, which explicitly incorporates the longest-range interaction from the OPE. We can, therefore, extract infinite volume observables from finite volume energy levels both below and above the lhc. Our method also
naturally accounts for range effects and exponentially suppressed corrections related to the OPE~\cite{Sato:2007ms}.
Additionally, the formulation of chiral EFT using the
plane wave basis~\cite{Meng:2021uhz}  enables us to investigate and understand the impact of partial-wave mixing in a finite volume on the extracted phase shifts.

In parallel to our work, a modified L\"uscher formula was proposed to address the lhc problem in Ref.~\cite{Raposo:2023oru}; however, no practical implementations of this approach to lattice data were conducted. 
To demonstrate our method, we undertake a thorough analysis of lattice energy levels from Ref.~\cite{Padmanath:2022cvl} to extract, for the first time, the pole of the $T_{cc}$ state while considering all the effects above and quantifying the uncertainties.
Our method is general and applicable to the analysis of a wide range of hadronic reactions utilizing lattice energy levels.

\section{Framework}
In the 1990s, Lüscher established  a method that connects  the infinite-volume scattering matrix $T(E)$ to the discrete energy levels $E_{FV}$ of a system in a periodic box~\cite{Luscher:1986pf,Luscher:1990ux}. 
The Lüscher formula, also known as Lüscher's quantization conditions (LQCs) can be schematically expressed as~\cite{Bulava:2022ovd,Briceno:2017max}
\be\label{eq:LuscherQC}
{ \rm det} [F^{-1}(E, {\bf  P}, L) -8 \pi i T(E)] =0,
\ee
where $F^{-1}(E, {\bf  P}, L)$ is a known  
quantity  
that captures the kinematics of the finite volume. It depends on the box size $L$, the total momentum of the two-body system $\bf P$ and 
the total energy $E$.
 Equation~\eqref{eq:LuscherQC}  determines a set of lattice energy levels $E_{FV}$ if the infinite-volume scattering amplitude $T(E)$ is known.
However, to obtain  observables in the infinite volume, a solution of the inverse problem is required, 
 where $T(E)$ is extracted  from $E_{FV}$.  The method is  applicable to two-body scattering, including  various   partial waves and  coupled hadron-hadron channels below  the
 lowest three-body threshold. 
However, while this approach is generally model-independent, it is 
valid under certain conditions.  First, the box size $L$ is required to be significantly larger than the interaction range $R$, in order to justify the neglect of exponentially suppressed corrections $\sim e^{-L/R} $. Yet, for not very large volumes, these exponentially suppressed terms, governed by the longest-range OPE interaction, can be numerically significant~\cite{Sato:2007ms}. Another complication stems from breaking the rotational symmetry in cubic boxes, which results in energy levels typically receiving contributions from multiple partial waves~\cite{Luscher:1990ux,Leskovec:2012gb}. Partial wave mixing is sometimes disregarded at very low energies due to the threshold suppression of higher partial waves, $T_l \sim E^l$, 
ensuring a one-to-one correspondence between the phase shifts and the FV energy levels. 
However, for more general cases where the partial wave mixing effect is significant, this correspondence is lost, and a more complex formalism involving appropriate parameterization of the $T$-matrix is required to determine scattering information,  see, e.g.,~\cite{Dudek:2012gj,Woss:2018irj} and references therein.
One option for parameterization is the ERE~\cite{Morningstar:2017spu}, which, however, is only valid in a very narrow energy range limited by the  lhc~\cite{Du:2023hlu}. 
Finally and most importantly,
Lüscher's quantization conditions fail in the presence of the nearby lhc \cite{Suppl_Tcc_EFV,Raposo:2023oru,Dawid:2023jrj}. Indeed, because the amplitude $T(E)$ is complex below it, while the function $F^{-1}(E,\bm{P},L)$ remains real, Eq.~\eqref{eq:LuscherQC} can not be applied at least below the lhc.

In this work, we advocate an alternative approach  (see also~\cite{Meng:2021uhz}), which allows one to 
account for all effects discussed above, thereby avoiding the complexity  of solving the inverse problem, see  Fig.~\ref{fig:our_method} for a schematic illustration. 
Specifically, we start from the effective Hamiltonian, which incorporates the long-range dynamics due to the OPE and involves contact interactions in relevant partial waves. We then calculate the FV energy spectrum using the plane wave basis with discrete momentum modes and adjust the 
 low-energy constants (LECs), accompanying the contact terms, to achieve the best description of the FV energy levels $E_{FV}$.   
  The resulting effective Hamiltonian, with all the LECs being fixed to $E_{FV}$, is then used to calculate the scattering amplitude in the infinite volume.

 \section{Application to $T_{cc}$}
\begin{figure}[t]
\begin{center}
\includegraphics[width=0.5\textwidth]{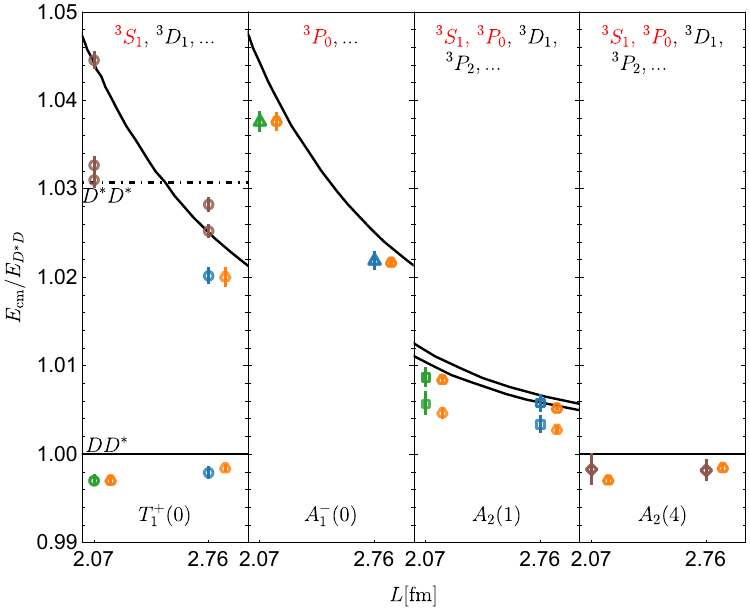}
\caption{ \label{fig:E_FV}
Fit results for the center-of-mass energy  $E_{\rm cm} =\sqrt{E^2 - {\bf P}^2 }$   
of the $DD^*$ system normalized by $E_{DD^*} = M_D + M_{D^*}$, for the heavier charm quark mass and two different volumes from Ref.~\cite{Padmanath:2022cvl}  in various FV irreps.  The lattice energy levels are shown by open circles, squares and  triangles: the blue and green points in the irreps $T_1^+(0), A_1^-(0)$ and $A_2(1)$ are used as input in this analysis as well as in the scattering analysis of Ref.~\cite{Padmanath:2022cvl}.  
  The orange symbols,   slightly shifted to the right for transparency, represent the results of our full calculation (Fit 2), including pions.  For each irrep, we indicate the lowest partial waves, which contribute to it. 
  Our results in the irrep  $A_2(4)$ are predictions. 
The solid  and dot-dashed lines correspond to  the noninteracting $DD^*$ and $D^*D^*$ energies, respectively.}
\end{center}
\end{figure}

In Ref.~\cite{Padmanath:2022cvl},  the FV energy levels of isospin-0 $DD^*$ scattering were extracted in lattice QCD using the lattice spacing of $a\approx$  0.08636 fm at $m_{\pi} \approx 280 $ MeV, corresponding to the 
   $D$ and  $D^*$ meson masses of $M_D =1927$ MeV and $M_{D^*}=2049$ MeV, respectively, 
  and two spatial lattice sizes $L=2.07$ and $2.76$ fm, as shown in Fig~\ref{fig:E_FV}. 
 Following Ref.~\cite{Padmanath:2022cvl}, we consider the nine lowest-lying energy levels in the irreducible representations (irreps) $T_1^+(0), A_1^-(0)$ and $A_2(1)$ of the point groups   as input in our calculations. The integer numbers $d$ in the parentheses are related to  the total momentum of two particles $\bm{P}={2\pi \over L} \bm d$ with $\bm d \in Z^3$. 
   
  Starting from the Lippmann-Schwinger-type integral equations  (LSE) in the FV, $\mathbb{T}(E)=\mathbb{V}(E)+\mathbb{V}(E)\mathbb{G}(E)\mathbb{T}(E),$
  the FV energy levels are obtained by solving
the determinant equation
\begin{equation}
	\det\left[\mathbb{G}^{-1}(E)-\mathbb{V}(E)\right]=0\label{eq:rel_det0}.
\end{equation}
 The matrix in the argument of the determinant can be block-diagonalized according to the lattice irreps. 
  Here, the discretized propagator $\mathbb{G}$ is defined as  
\bea
	\mathbb{G}_{\bm{n},\bm{n}'}={\cal J}\frac{1}{L^{3}}G(\tilde p_{\bm{n}},E)\delta_{\bm{n}',\bm{n}},
\eea
 where $\mathcal{J}$ is the Jacobi determinant arising from the transformation between the box  and the center-of-mass  frames,  see  \cite{Suppl_Tcc_EFV} for details,  
 while $\tilde p_{\bm{n}}$ are the discretized momenta. 
Further, the Green function $G$ reads 
\bea
G(\tilde p,E)=\frac{1}{4\omega_{1}\omega_{2}} \left(  \frac1{E-\omega_1-\omega_2} - \frac1{E +\omega_1+\omega_2} \right),
\eea
where $\omega_i = \sqrt{m_i^2+\tilde p^2}$  with $m_1 =M_D$  and $m_2=M_{D^*}$.   
To solve Eq.~\eqref{eq:rel_det0} in a finite volume, we use the plane wave basis instead of expanding it in partial waves. 
This allows us to naturally account for all partial wave mixing effects arising from rotational symmetry breaking in a cubic box~\cite{Meng:2021uhz}.

The effective potential $V$ is constructed in chiral   EFT up to ${\cal O}(Q^2)$, with $Q\sim m_{\pi}$ being the soft scale of the expansion,  and reads
\begin{equation}
V=V_{\text{OPE}}^{(0)}+V_{\text{cont}}^{(0)}+V_{\text{cont}}^{(2)}+...,\label{eq:veft}
\end{equation}
where the two-pion exchange contributions at the considered value of $m_\pi$ are assumed to be saturated by the contact terms.
 Truncating the contact interactions to ${\cal O}(Q^2)$, the most relevant contact potentials contributing to the irreps $T_1^+(0), A_1^-(0)$ and $A_2(1)$ read  
\bea\nonumber
V_{\text{cont}}^{(0)+(2)}[^3S_1]&=&\left(C^{(0)}_{^{3}S_{1}}  +C^{(2)}_{^{3}S_{1}}  (p^{2}+p'^{2})\right ) ({\bm\epsilon \cdot \bm\epsilon'^*}  ) \\
V_{\text{cont}}^{(2)}[^3P_0]&=&C^{(2)}_{^{3}P_{0}}  ({\bm p'\cdot \bm\epsilon'^*}  )  ({\bm p \cdot \bm\epsilon  }  )     ~\label{eq:Vct},
\eea
where $\bm p$ $(\bm p')$ and $\bm\epsilon$ $(\bm\epsilon')$ denote the cms momentum  and   polarization of the initial (final) $D^*$ meson, respectively.
While the irreps $T_1^+(0)$ and $A_2(1)$ can also receive contributions from the 
$S$-to-$D$-wave short-range interactions as well as from the $^3P_2$ partial waves at ${\cal O}(Q^2)$,  in what follows, we 
consider  fits with 3-parameters from  Eq.~\eqref{eq:Vct} as our main results and  use the additional contributions from other partial waves to estimate systematic uncertainties in \cite{Suppl_Tcc_EFV}. The longest-range interaction between the $D$ and $D^*$ mesons is driven by the OPE, which in the static approximation reads
\be\label{Eq:OPE}
V_{\text{OPE}}^{(0)} =  -3\frac{M_{D}M_{D^{*}}g^2}{f_{\pi}^2} \; \frac{ ({\bm k\cdot \bm\epsilon}  )  ({\bm k \cdot \bm\epsilon'^*  }) }{\bm k^2 + \mu^2},
\ee
where $ \mu^2=m_{\pi}^2-\Delta M^2$,  $\Delta M = M_{D^*}-M_D$ and $\bm k=\bm p'+\bm p$. The pion mass dependence of the pion decay constant $f_{\pi}$  is considered along the lines of Ref.~\cite{Du:2023hlu,Becirevic:2012pf},
which gives $f_{\pi}=105.3$ MeV for $m_{\pi}=280$ MeV. The value of the coupling constant $g$ is extracted from the fits  to its physical value and the lattice data of Ref.~\cite{Becirevic:2012pf}. For the given lattice spacing of $a\approx 0.086$ fm and $m_{\pi}=280$ MeV, we found $g=0.517\pm 0.015$ \cite{Suppl_Tcc_EFV}.
When both  $p$ and $ p'$ are on shell,  $ p=p'=\frac{\lambda(E^{2},M_{D}^{2},M_{D^{*}}^{2})^{1/2}}{2E}$ ($\lambda$ is the Källén function), the OPE and, consequently,  the on-shell $DD^*$ partial wave amplitudes, exhibit the lhc with the closest to the threshold branch point given by \cite{Du:2023hlu}
\be
(p_\lhc^{1\pi})^2=-\frac{\mu^2}4=-(126 \ \mbox{MeV})^2\Rightarrow \left(\frac{p_\lhc^{1\pi}}{E_{DD^*}}\right)^2\approx-0.001, ~~\label{eq:lhc_num}
\ee
where $E_{DD^*}=M_D+M_{D^*}$. In principle, the OPE may also have the  three-body right-hand cut, corresponding to the on shell $DD\pi$  state.  However, for $m_\pi = 280$~MeV, it starts at momenta far away from the threshold, $p_{\rhc_3}^2=(552 \ \mbox{MeV})^2$ \cite{Du:2023hlu},
which makes it irrelevant for the current analysis.  It should be noticed that all partial waves  are included in Eq.~\eqref{Eq:OPE}, as no partial wave expansion and truncation is made for the OPE  in our plane wave expansion method.

The contact interactions in the LSE are supplemented  with the exponential regulators of the form $e^{\frac{-(p^n+p'^n)}{\Lambda^n}}$  with $n=6$. 
The regularization of the operators with the single pion propagator preserving long-range dynamics is worked out in Ref.~\cite{Reinert:2017usi} and can be implemented by a substitution:
\be
\frac1{\bm k^2 + \mu^2}\to \frac1{\bm k^2 + \mu^2} e^{\frac{-(\bm k^2+\mu^2)}{\Lambda^2}}.
\ee
In what follows,  
we present the results for  the cutoff $\Lambda=0.9$ GeV  
and consider the cutoff variation from  0.7 to 1.2 GeV to estimate systematic uncertainties in \cite{Suppl_Tcc_EFV}. 

\begin{figure*}[t]
\begin{center}
\includegraphics[width=0.9\textwidth]{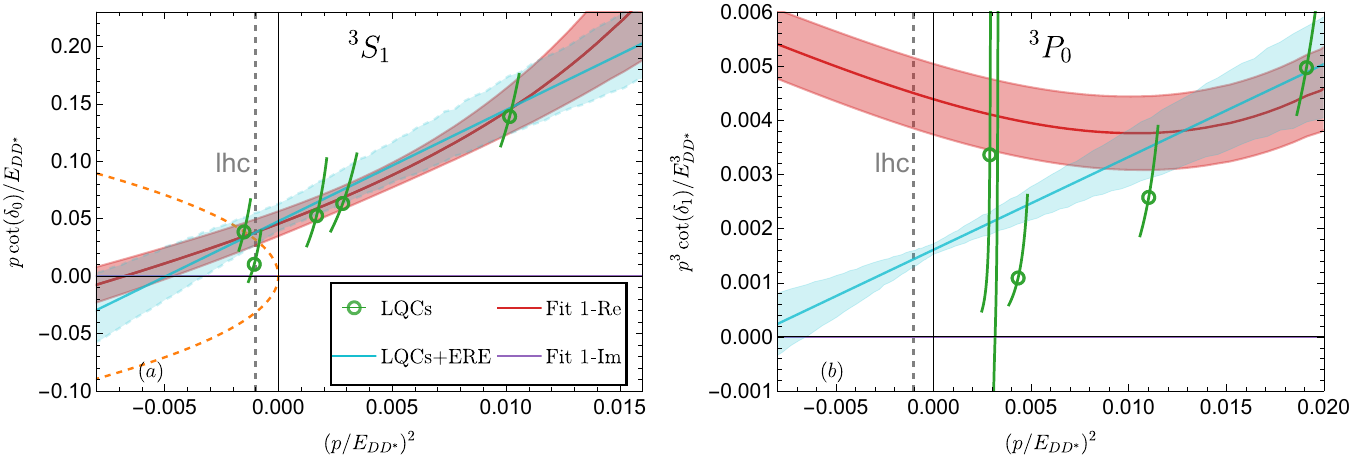}
\includegraphics[width=0.9\textwidth]{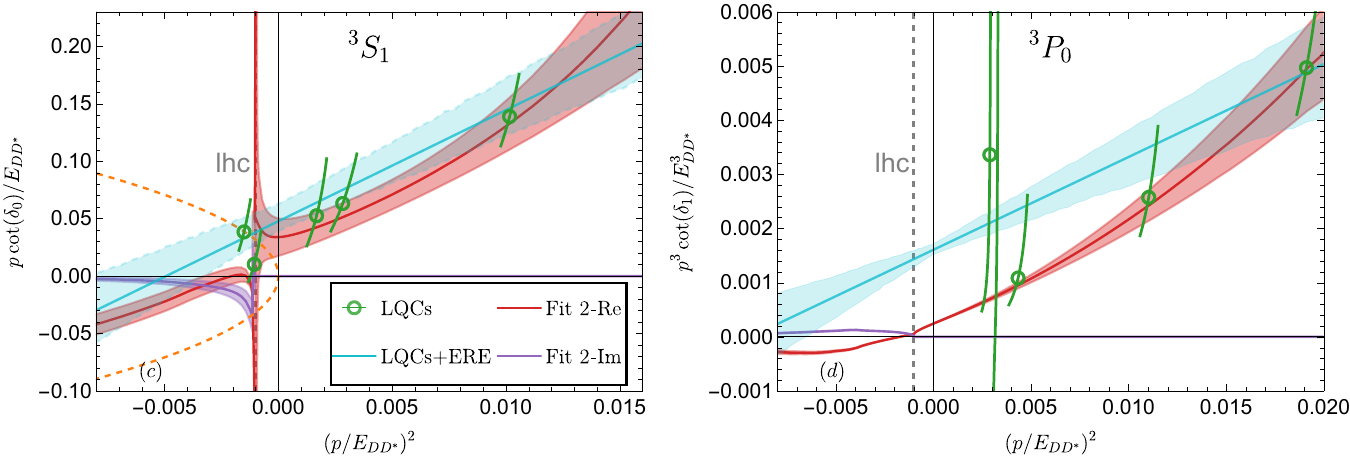}
\caption{\label{fig:phase_shifts}  
Phase shifts in the $^3S_1$ (left panel) and $^3P_0$ (right panel) partial waves extracted from the FV energy levels $E_{FV}$ calculated in lattice QCD. Red bands represent the results of our 3-parameter fits to  $E_{FV}$
\emph{without} the OPE (Fit 1, upper panel) and \emph{with} the OPE  (Fit 2, lower panel), including the $1\sigma$ uncertainty.
   Green dots in the left panel are the  
 phase shifts extracted from $E_{FV}$ using the single-channel Lüscher quantization conditions in Ref.~\cite{Padmanath:2022cvl}.
 Green dots in the right panel are extracted in this study using the same method.
Blue bands are the results of the 4-parameter fits of $E_{FV}$ using the ERE  in   Ref.~\cite{Padmanath:2022cvl}. Orange lines in the left panel correspond to 
$ip=\pm |p|$ from unitarity,   normalized to $E_{DD^*}$. The gray vertical dashed line denotes the position of the branch point of the left-hand cut nearest to the threshold. } 
\end{center}
\end{figure*}
To see the impact of the OPE on the results, we  perform two calculations: In Fit 1, we start from a pure contact potential without the OPE
and adjust the LECs  $C^{(0)}_{^{3}S_{1}}, C^{(2)}_{^{3}S_{1}}$ and $C^{(2)}_{^{3}P_{0}}$  to obtain the best $\chi^2$ fit to $E_{FV}$.  In Fit 2, we include, in addition to  the contact interactions, also the OPE; the corresponding results for the energy levels are shown in  Fig.~\ref{fig:E_FV}. For both fits partial wave mixing is included when calculating $E_{FV}$ in different lattice irreps. The OPE, however, induces additional mixing between $S$ and $D$ waves due to the long-range tensor interactions. Furthermore, the OPE introduces a new momentum scale related with the branch point of the lhc in Eq.~\eqref{eq:lhc_num}, which has several important consequences on the observables:  (i) It modifies the analytic structure of the scattering amplitude, making, in particular, the phase shifts complex when analytically continued below the lhc;  (ii)  It controls the energy dependence of the scattering amplitude in the near-threshold region and (iii)  It governs the leading exponentially  suppressed corrections $\sim e^{-\mu  L}$, neglected in the Lüscher approach.  
 
 \begin{table*}[htp]
     \centering
\begin{tabular*}{\hsize}{@{}@{\extracolsep{\fill}}lcccccc@{}}
\hline 
\hline 
 & $a_{^{3}\!S_{1}}$ {[}fm{]} & $r_{^{3}\!S_{1}}$ {[}fm{]} & $\delta m_{T_{cc}}$ {[}MeV{]}  & $a_{^{3}\!P_{0}}$ {[}fm$^{3}${]} & $r_{^{3}\!P_{0}}$ {[}fm$^{-1}${]} & $\chi^{2}/\text{dof}$\tabularnewline
\hline 

LQCs+ERE fit~\cite{Padmanath:2022cvl} & $1.04\pm0.29$ & 0.96$_{-0.20}^{+0.18}$ & $-9.9_{-7.2}^{+3.6}$  & $0.076_{-0.009}^{+0.008}$ & $6.9\pm2.1$ & $3.7/5$\tabularnewline
 Fit 1: cont. & $1.09\pm0.35$ & $0.75\pm0.14$ & $-10.6\pm4.4$   & $0.028\pm0.004$ & $-4.3\pm0.05$ & $5.52/6$\tabularnewline
Fit 2: cont.+OPE & $1.46\pm0.57$ & $0.096\pm0.53$ & $-6.6(\pm1.5)-i4.0(\pm3.7)$ & $0.497\pm0.007$ & $5.63\pm0.19$ & $2.95/6$\tabularnewline
\hline
\hline 
\end{tabular*}
     \caption{\label{tab:ere_pole_num}
     Results for the $T_{cc}$ pole position $\delta m_{T_{cc}}$,  defined relative to the $DD^*$ threshold, and the $DD^*$ ERE parameters.
     The results obtained via the Lüscher QCs plus ERE in Ref.~\cite{Padmanath:2022cvl} are present in the second row. Our results from  Fits 1 and 2  
     are given in the third and forth rows, respectively. 
     The   $T_{cc}$ pole position in Fit 2 corresponds to a resonance state with 85\% probability within the $1\sigma$ uncertainty. The residual 
     15\% probability corresponds to a scenario with two virtual poles--see the intersection area of the red band with the orange curve in Fig.~\ref{fig:phase_shifts}. 
 }  
 \end{table*}
 
 With the LECs fixed from the best fits to $E_{FV}$, we are in the position to calculate the infinite volume observables and confront them with the results of the Lüscher 
 analysis of Ref.~\cite{Padmanath:2022cvl}. 
 In Fig.~\ref{fig:phase_shifts}, the results are shown for the phase shifts in the $^3S_1$ and $^3P_0$ partial waves. 
 In the vicinity of the threshold, the phase shifts can be expanded 
 employing the ERE
 \begin{equation}   \label{eq:ERE} p^{2l+1}\cot\delta^{(l,J)}=\frac{1}{a^{(l,J)}}+\frac{1}{2}r^{(l,J)}p^{2}+\ldots\, .
 \end{equation}
The predictions of Fit 1 for  $\delta_{^3\! S_1}$ (upper left panel) are consistent with the  analysis of Ref.~\cite{Padmanath:2022cvl} using the ERE (\ref{eq:ERE}) and also yield very similar values for the ERE parameters and the pole position of the $T_{cc}$ state, as summarized in Table \ref{tab:ere_pole_num}. This is not surprising since both analyses involve two parameters in this partial wave, which can be matched to the scattering length and effective range. On the other hand, the contact fit results for the $\delta_{^3\! P_0}$ 
are unable to describe all the data points since the  low-energy behavior of the phase shifts can not be captured with a single-parameter fit. To account for the range corrections, 
the two-parameter fit was introduced  in Ref.~\cite{Padmanath:2022cvl} in line with Eq.~\eqref{eq:ERE}.  
This is, however, not needed, since the range corrections in this channel are almost completely driven by the OPE -- see our Fit 2 in the lower right panel. The effect of the OPE on $\delta_{^3\! S_1}$  is also very substantial. The nontrivial interplay of the repulsive OPE and attractive short range interactions results in the appearance of a pole in $p \cot \delta_{^3\! S_1}$ in the vicinity of the lhc, in line with the results of Ref.~\cite{Du:2023hlu}. 
 This significantly impacts the validity range of the ERE, the extracted values of the scattering length and effective range, as well as the $T_{cc}$ pole position, which, in our
  calculation, is highly likely to be a resonance state -- see Table~\ref{tab:ere_pole_num} for details.
 In addition, comparing the phase shifts extracted using the Lüscher approach (green points) with our Fit 2 reveals discrepancies, 
 in particular,  for the two lowest-energy datapoints, which are strongly influenced by the lhc. 
 On the other hand, the Lüscher method is consistent with our analysis above the $DD^*$ threshold for both $\delta_{^3\! S_1}$ and $\delta_{^3\! P_0}$ phase shifts within errors.

 \section{Summary and outlook}
We discuss a novel approach based on effective field theory to extract information on two-body scattering from finite-volume energies { relying on the chiral expansion  at low energies}. Its main advantage as compared to
the  L\" uscher method consists in the explicit account for the longest-range interaction including the leading left-hand cut, which is crucial for maintaining the appropriate analytic structure of the scattering amplitude near the threshold.  Using this method, the finite-volume energy levels can be directly calculated  as solutions of the eigenvalue problem both below and above the left-hand cut.
It also addresses range effects and the leading exponentially suppressed corrections from the longest-range interaction in a model-independent way.  The efficacy of our calculation benefits from using the plane wave basis expansion and the eigenvector continuation -- the modern computational technique to fully incorporate the partial-wave mixing effects on lattice and to effectively solve the eigenvalue problem with the small computational cost. 

  The practical advantages of the approach  are demonstrated by making a comprehensive analysis of  the lattice energy levels  on $DD^*$ scattering from~\cite{Padmanath:2022cvl} in connection to  the doubly charm tetraquark, understanding the properties of which is of fundamental importance in the context of the XYZ exotic states. 
The long-range interaction from the OPE
is demonstrated to significantly influence the understanding of infinite volume observables. Its presence governs the range effects in the $^3P_0$ channel and,   contrary to  the L\" uscher method, allows one to properly calculate amplitudes  in the vicinity of the left-hand cut.
 The systematic corrections related to the truncation of the EFT expansion are shown to be small compared to statistical uncertainties. 
The extracted pole position of the $T_{cc}^+$ state appears to be most likely a below-threshold resonance shifted to the complex plane due to the OPE. If the uncertainty of the energy levels is substantially reduced, our approach can be used to directly extract the strength of the OPE, represented by the ratio $g/f_{\pi}$,  from lattice data. 
 The incorporation of three-body ($DD\pi$)   right-hand cuts  is also straightforward and expected to play an important role for analyzing lattice data for lower values of the pion masses  -- see also~\cite{Hansen:2024ffk}.

Our approach is applicable to a wide range of hadronic systems at unphysical pion masses, where finite volume energy levels are already available or will be computed in lattice simulations. For instance, it can enhance our understanding of nucleon-nucleon scattering, where partial wave mixing effects are expected to be important at the physical values of the quark masses { and the effects from the lhc are significant} \cite{Meng:2021uhz},
 and can shed light on hyperon-nucleon and hyperon-hyperon scattering, difficult to obtain otherwise. 
Consequently, the effect of the lhc is expected to be relevant for  probing nuclear structure,  neutron stars, $\Sigma$ hypernuclei, D-mesic nuclei, etc.,  but also for understanding exotic hadrons -- tetraquarks \cite{Prelovsek:2013cra,Padmanath:2022cvl}, pentaquarks \cite{Bicudo:2022cqi} and even six-quark states \cite{Green:2021qol}.  
These investigations provide important insights into QCD dynamics and its manifestations in the hadron spectrum and reactions.

\begin{acknowledgments}

The authors are grateful to John Bulava, Jambul Gegelia  and Christoph Hanhart  for sharing their insights into the considered topics and to Sasa Prelovsek and  Madanagopalan
Padmanath  for providing us with the covariance matrix  and for valuable comments to the manuscript. 
L.M. is grateful to the helpful discussions with Yan Li and Zi-Yang Lin. 
   This work is supported in part BMBF (contract No. 05P21PCFP1), by DFG and NSFC through
funds provided to the Sino-German CRC 110 ``Symmetries and the
Emergence of Structure in QCD'' (NSFC Grant No. 11621131001, DFG Grant
No. TRR110), by the MKW NRW under the funding code NW21-024-A and by
the EU Horizon 2020 research and innovation programme (STRONG-2020,
grant agreement No. 824093) and by the European Research Council (ERC) under the EU
Horizon 2020 research and innovation programme (ERC AdG
NuclearTheory, grant agreement No. 885150).

\end{acknowledgments}

\bibliography{Tcc_refs}

\begin{thebibliography}{65}%
\makeatletter
\providecommand \@ifxundefined [1]{%
 \@ifx{#1\undefined}
}%
\providecommand \@ifnum [1]{%
 \ifnum #1\expandafter \@firstoftwo
 \else \expandafter \@secondoftwo
 \fi
}%
\providecommand \@ifx [1]{%
 \ifx #1\expandafter \@firstoftwo
 \else \expandafter \@secondoftwo
 \fi
}%
\providecommand \natexlab [1]{#1}%
\providecommand \enquote  [1]{``#1''}%
\providecommand \bibnamefont  [1]{#1}%
\providecommand \bibfnamefont [1]{#1}%
\providecommand \citenamefont [1]{#1}%
\providecommand \href@noop [0]{\@secondoftwo}%
\providecommand \href [0]{\begingroup \@sanitize@url \@href}%
\providecommand \@href[1]{\@@startlink{#1}\@@href}%
\providecommand \@@href[1]{\endgroup#1\@@endlink}%
\providecommand \@sanitize@url [0]{\catcode `\\12\catcode `\$12\catcode `\&12\catcode `\#12\catcode `\^12\catcode `\_12\catcode `\%12\relax}%
\providecommand \@@startlink[1]{}%
\providecommand \@@endlink[0]{}%
\providecommand \url  [0]{\begingroup\@sanitize@url \@url }%
\providecommand \@url [1]{\endgroup\@href {#1}{\urlprefix }}%
\providecommand \urlprefix  [0]{URL }%
\providecommand \Eprint [0]{\href }%
\providecommand \doibase [0]{https://doi.org/}%
\providecommand \selectlanguage [0]{\@gobble}%
\providecommand \bibinfo  [0]{\@secondoftwo}%
\providecommand \bibfield  [0]{\@secondoftwo}%
\providecommand \translation [1]{[#1]}%
\providecommand \BibitemOpen [0]{}%
\providecommand \bibitemStop [0]{}%
\providecommand \bibitemNoStop [0]{.\EOS\space}%
\providecommand \EOS [0]{\spacefactor3000\relax}%
\providecommand \BibitemShut  [1]{\csname bibitem#1\endcsname}%
\let\auto@bib@innerbib\@empty
\bibitem [{\citenamefont {Esposito}\ \emph {et~al.}(2017)\citenamefont {Esposito}, \citenamefont {Pilloni},\ and\ \citenamefont {Polosa}}]{Esposito:2016noz}%
  \BibitemOpen
  \bibfield  {author} {\bibinfo {author} {\bibfnamefont {A.}~\bibnamefont {Esposito}}, \bibinfo {author} {\bibfnamefont {A.}~\bibnamefont {Pilloni}},\ and\ \bibinfo {author} {\bibfnamefont {A.~D.}\ \bibnamefont {Polosa}},\ }\bibfield  {title} {\bibinfo {title} {{Multiquark Resonances}},\ }\href {https://doi.org/10.1016/j.physrep.2016.11.002} {\bibfield  {journal} {\bibinfo  {journal} {Phys. Rept.}\ }\textbf {\bibinfo {volume} {668}},\ \bibinfo {pages} {1} (\bibinfo {year} {2017})},\ \Eprint {https://arxiv.org/abs/1611.07920} {arXiv:1611.07920 [hep-ph]} \BibitemShut {NoStop}%
\bibitem [{\citenamefont {Lebed}\ \emph {et~al.}(2017)\citenamefont {Lebed}, \citenamefont {Mitchell},\ and\ \citenamefont {Swanson}}]{Lebed:2016hpi}%
  \BibitemOpen
  \bibfield  {author} {\bibinfo {author} {\bibfnamefont {R.~F.}\ \bibnamefont {Lebed}}, \bibinfo {author} {\bibfnamefont {R.~E.}\ \bibnamefont {Mitchell}},\ and\ \bibinfo {author} {\bibfnamefont {E.~S.}\ \bibnamefont {Swanson}},\ }\bibfield  {title} {\bibinfo {title} {{Heavy-Quark QCD Exotica}},\ }\href {https://doi.org/10.1016/j.ppnp.2016.11.003} {\bibfield  {journal} {\bibinfo  {journal} {Prog. Part. Nucl. Phys.}\ }\textbf {\bibinfo {volume} {93}},\ \bibinfo {pages} {143} (\bibinfo {year} {2017})},\ \Eprint {https://arxiv.org/abs/1610.04528} {arXiv:1610.04528 [hep-ph]} \BibitemShut {NoStop}%
\bibitem [{\citenamefont {Guo}\ \emph {et~al.}(2018)\citenamefont {Guo}, \citenamefont {Hanhart}, \citenamefont {Mei\ss{}ner}, \citenamefont {Wang}, \citenamefont {Zhao},\ and\ \citenamefont {Zou}}]{Guo:2017jvc}%
  \BibitemOpen
  \bibfield  {author} {\bibinfo {author} {\bibfnamefont {F.-K.}\ \bibnamefont {Guo}}, \bibinfo {author} {\bibfnamefont {C.}~\bibnamefont {Hanhart}}, \bibinfo {author} {\bibfnamefont {U.-G.}\ \bibnamefont {Mei\ss{}ner}}, \bibinfo {author} {\bibfnamefont {Q.}~\bibnamefont {Wang}}, \bibinfo {author} {\bibfnamefont {Q.}~\bibnamefont {Zhao}},\ and\ \bibinfo {author} {\bibfnamefont {B.-S.}\ \bibnamefont {Zou}},\ }\bibfield  {title} {\bibinfo {title} {{Hadronic molecules}},\ }\href {https://doi.org/10.1103/RevModPhys.90.015004} {\bibfield  {journal} {\bibinfo  {journal} {Rev. Mod. Phys.}\ }\textbf {\bibinfo {volume} {90}},\ \bibinfo {pages} {015004} (\bibinfo {year} {2018})},\ \Eprint {https://arxiv.org/abs/1705.00141} {arXiv:1705.00141 [hep-ph]} \BibitemShut {NoStop}%
\bibitem [{\citenamefont {Yamaguchi}\ \emph {et~al.}(2020)\citenamefont {Yamaguchi}, \citenamefont {Hosaka}, \citenamefont {Takeuchi},\ and\ \citenamefont {Takizawa}}]{Yamaguchi:2019vea}%
  \BibitemOpen
  \bibfield  {author} {\bibinfo {author} {\bibfnamefont {Y.}~\bibnamefont {Yamaguchi}}, \bibinfo {author} {\bibfnamefont {A.}~\bibnamefont {Hosaka}}, \bibinfo {author} {\bibfnamefont {S.}~\bibnamefont {Takeuchi}},\ and\ \bibinfo {author} {\bibfnamefont {M.}~\bibnamefont {Takizawa}},\ }\bibfield  {title} {\bibinfo {title} {{Heavy hadronic molecules with pion exchange and quark core couplings: a guide for practitioners}},\ }\href {https://doi.org/10.1088/1361-6471/ab72b0} {\bibfield  {journal} {\bibinfo  {journal} {J. Phys. G}\ }\textbf {\bibinfo {volume} {47}},\ \bibinfo {pages} {053001} (\bibinfo {year} {2020})},\ \Eprint {https://arxiv.org/abs/1908.08790} {arXiv:1908.08790 [hep-ph]} \BibitemShut {NoStop}%
\bibitem [{\citenamefont {Brambilla}\ \emph {et~al.}(2020)\citenamefont {Brambilla}, \citenamefont {Eidelman}, \citenamefont {Hanhart}, \citenamefont {Nefediev}, \citenamefont {Shen}, \citenamefont {Thomas}, \citenamefont {Vairo},\ and\ \citenamefont {Yuan}}]{Brambilla:2019esw}%
  \BibitemOpen
  \bibfield  {author} {\bibinfo {author} {\bibfnamefont {N.}~\bibnamefont {Brambilla}}, \bibinfo {author} {\bibfnamefont {S.}~\bibnamefont {Eidelman}}, \bibinfo {author} {\bibfnamefont {C.}~\bibnamefont {Hanhart}}, \bibinfo {author} {\bibfnamefont {A.}~\bibnamefont {Nefediev}}, \bibinfo {author} {\bibfnamefont {C.-P.}\ \bibnamefont {Shen}}, \bibinfo {author} {\bibfnamefont {C.~E.}\ \bibnamefont {Thomas}}, \bibinfo {author} {\bibfnamefont {A.}~\bibnamefont {Vairo}},\ and\ \bibinfo {author} {\bibfnamefont {C.-Z.}\ \bibnamefont {Yuan}},\ }\bibfield  {title} {\bibinfo {title} {{The $XYZ$ states: experimental and theoretical status and perspectives}},\ }\href {https://doi.org/10.1016/j.physrep.2020.05.001} {\bibfield  {journal} {\bibinfo  {journal} {Phys. Rept.}\ }\textbf {\bibinfo {volume} {873}},\ \bibinfo {pages} {1} (\bibinfo {year} {2020})},\ \Eprint {https://arxiv.org/abs/1907.07583} {arXiv:1907.07583 [hep-ex]} \BibitemShut {NoStop}%
\bibitem [{\citenamefont {Guo}\ \emph {et~al.}(2020)\citenamefont {Guo}, \citenamefont {Liu},\ and\ \citenamefont {Sakai}}]{Guo:2019twa}%
  \BibitemOpen
  \bibfield  {author} {\bibinfo {author} {\bibfnamefont {F.-K.}\ \bibnamefont {Guo}}, \bibinfo {author} {\bibfnamefont {X.-H.}\ \bibnamefont {Liu}},\ and\ \bibinfo {author} {\bibfnamefont {S.}~\bibnamefont {Sakai}},\ }\bibfield  {title} {\bibinfo {title} {{Threshold cusps and triangle singularities in hadronic reactions}},\ }\href {https://doi.org/10.1016/j.ppnp.2020.103757} {\bibfield  {journal} {\bibinfo  {journal} {Prog. Part. Nucl. Phys.}\ }\textbf {\bibinfo {volume} {112}},\ \bibinfo {pages} {103757} (\bibinfo {year} {2020})},\ \Eprint {https://arxiv.org/abs/1912.07030} {arXiv:1912.07030 [hep-ph]} \BibitemShut {NoStop}%
\bibitem [{\citenamefont {Chen}\ \emph {et~al.}(2023)\citenamefont {Chen}, \citenamefont {Chen}, \citenamefont {Liu}, \citenamefont {Liu},\ and\ \citenamefont {Zhu}}]{Chen:2022asf}%
  \BibitemOpen
  \bibfield  {author} {\bibinfo {author} {\bibfnamefont {H.-X.}\ \bibnamefont {Chen}}, \bibinfo {author} {\bibfnamefont {W.}~\bibnamefont {Chen}}, \bibinfo {author} {\bibfnamefont {X.}~\bibnamefont {Liu}}, \bibinfo {author} {\bibfnamefont {Y.-R.}\ \bibnamefont {Liu}},\ and\ \bibinfo {author} {\bibfnamefont {S.-L.}\ \bibnamefont {Zhu}},\ }\bibfield  {title} {\bibinfo {title} {{An updated review of the new hadron states}},\ }\href {https://doi.org/10.1088/1361-6633/aca3b6} {\bibfield  {journal} {\bibinfo  {journal} {Rept. Prog. Phys.}\ }\textbf {\bibinfo {volume} {86}},\ \bibinfo {pages} {026201} (\bibinfo {year} {2023})},\ \Eprint {https://arxiv.org/abs/2204.02649} {arXiv:2204.02649 [hep-ph]} \BibitemShut {NoStop}%
\bibitem [{\citenamefont {Meng}\ \emph {et~al.}(2023{\natexlab{a}})\citenamefont {Meng}, \citenamefont {Wang}, \citenamefont {Wang},\ and\ \citenamefont {Zhu}}]{Meng:2022ozq}%
  \BibitemOpen
  \bibfield  {author} {\bibinfo {author} {\bibfnamefont {L.}~\bibnamefont {Meng}}, \bibinfo {author} {\bibfnamefont {B.}~\bibnamefont {Wang}}, \bibinfo {author} {\bibfnamefont {G.-J.}\ \bibnamefont {Wang}},\ and\ \bibinfo {author} {\bibfnamefont {S.-L.}\ \bibnamefont {Zhu}},\ }\bibfield  {title} {\bibinfo {title} {{Chiral perturbation theory for heavy hadrons and chiral effective field theory for heavy hadronic molecules}},\ }\href {https://doi.org/10.1016/j.physrep.2023.04.003} {\bibfield  {journal} {\bibinfo  {journal} {Phys. Rept.}\ }\textbf {\bibinfo {volume} {1019}},\ \bibinfo {pages} {1} (\bibinfo {year} {2023}{\natexlab{a}})},\ \Eprint {https://arxiv.org/abs/2204.08716} {arXiv:2204.08716 [hep-ph]} \BibitemShut {NoStop}%
\bibitem [{\citenamefont {Briceno}\ \emph {et~al.}(2018)\citenamefont {Briceno}, \citenamefont {Dudek},\ and\ \citenamefont {Young}}]{Briceno:2017max}%
  \BibitemOpen
  \bibfield  {author} {\bibinfo {author} {\bibfnamefont {R.~A.}\ \bibnamefont {Briceno}}, \bibinfo {author} {\bibfnamefont {J.~J.}\ \bibnamefont {Dudek}},\ and\ \bibinfo {author} {\bibfnamefont {R.~D.}\ \bibnamefont {Young}},\ }\bibfield  {title} {\bibinfo {title} {{Scattering processes and resonances from lattice QCD}},\ }\href {https://doi.org/10.1103/RevModPhys.90.025001} {\bibfield  {journal} {\bibinfo  {journal} {Rev. Mod. Phys.}\ }\textbf {\bibinfo {volume} {90}},\ \bibinfo {pages} {025001} (\bibinfo {year} {2018})},\ \Eprint {https://arxiv.org/abs/1706.06223} {arXiv:1706.06223 [hep-lat]} \BibitemShut {NoStop}%
\bibitem [{\citenamefont {Aoki}\ and\ \citenamefont {Doi}(2020)}]{Aoki:2020bew}%
  \BibitemOpen
  \bibfield  {author} {\bibinfo {author} {\bibfnamefont {S.}~\bibnamefont {Aoki}}\ and\ \bibinfo {author} {\bibfnamefont {T.}~\bibnamefont {Doi}},\ }\bibfield  {title} {\bibinfo {title} {{Lattice QCD and baryon-baryon interactions: HAL QCD method}},\ }\href {https://doi.org/10.3389/fphy.2020.00307} {\bibfield  {journal} {\bibinfo  {journal} {Front. in Phys.}\ }\textbf {\bibinfo {volume} {8}},\ \bibinfo {pages} {307} (\bibinfo {year} {2020})},\ \Eprint {https://arxiv.org/abs/2003.10730} {arXiv:2003.10730 [hep-lat]} \BibitemShut {NoStop}%
\bibitem [{\citenamefont {Mai}\ \emph {et~al.}(2021)\citenamefont {Mai}, \citenamefont {D\"oring},\ and\ \citenamefont {Rusetsky}}]{Mai:2021lwb}%
  \BibitemOpen
  \bibfield  {author} {\bibinfo {author} {\bibfnamefont {M.}~\bibnamefont {Mai}}, \bibinfo {author} {\bibfnamefont {M.}~\bibnamefont {D\"oring}},\ and\ \bibinfo {author} {\bibfnamefont {A.}~\bibnamefont {Rusetsky}},\ }\bibfield  {title} {\bibinfo {title} {{Multi-particle systems on the lattice and chiral extrapolations: a brief review}},\ }\href {https://doi.org/10.1140/epjs/s11734-021-00146-5} {\bibfield  {journal} {\bibinfo  {journal} {Eur. Phys. J. ST}\ }\textbf {\bibinfo {volume} {230}},\ \bibinfo {pages} {1623} (\bibinfo {year} {2021})},\ \Eprint {https://arxiv.org/abs/2103.00577} {arXiv:2103.00577 [hep-lat]} \BibitemShut {NoStop}%
\bibitem [{\citenamefont {Bicudo}(2023)}]{Bicudo:2022cqi}%
  \BibitemOpen
  \bibfield  {author} {\bibinfo {author} {\bibfnamefont {P.}~\bibnamefont {Bicudo}},\ }\bibfield  {title} {\bibinfo {title} {{Tetraquarks and pentaquarks in lattice QCD with light and heavy quarks}},\ }\href {https://doi.org/10.1016/j.physrep.2023.10.001} {\bibfield  {journal} {\bibinfo  {journal} {Phys. Rept.}\ }\textbf {\bibinfo {volume} {1039}},\ \bibinfo {pages} {1} (\bibinfo {year} {2023})},\ \Eprint {https://arxiv.org/abs/2212.07793} {arXiv:2212.07793 [hep-lat]} \BibitemShut {NoStop}%
\bibitem [{\citenamefont {Bulava}\ \emph {et~al.}(2022)\citenamefont {Bulava} \emph {et~al.}}]{Bulava:2022ovd}%
  \BibitemOpen
  \bibfield  {author} {\bibinfo {author} {\bibfnamefont {J.}~\bibnamefont {Bulava}} \emph {et~al.},\ }\bibfield  {title} {\bibinfo {title} {{Hadron Spectroscopy with Lattice QCD}},\ }in\ \href@noop {} {\emph {\bibinfo {booktitle} {{Snowmass 2021}}}}\ (\bibinfo {year} {2022})\ \Eprint {https://arxiv.org/abs/2203.03230} {arXiv:2203.03230 [hep-lat]} \BibitemShut {NoStop}%
\bibitem [{\citenamefont {Prelovsek}(2023)}]{Prelovsek:2023sta}%
  \BibitemOpen
  \bibfield  {author} {\bibinfo {author} {\bibfnamefont {S.}~\bibnamefont {Prelovsek}},\ }\bibfield  {title} {\bibinfo {title} {{Spectroscopy of hadrons with heavy quarks from lattice QCD}},\ }\href@noop {} {\  (\bibinfo {year} {2023})},\ \Eprint {https://arxiv.org/abs/2310.07341} {arXiv:2310.07341 [hep-lat]} \BibitemShut {NoStop}%
\bibitem [{\citenamefont {Aaij}\ \emph {et~al.}(2022{\natexlab{a}})\citenamefont {Aaij} \emph {et~al.}}]{LHCb:2021vvq}%
  \BibitemOpen
  \bibfield  {author} {\bibinfo {author} {\bibfnamefont {R.}~\bibnamefont {Aaij}} \emph {et~al.} (\bibinfo {collaboration} {LHCb}),\ }\bibfield  {title} {\bibinfo {title} {{Observation of an exotic narrow doubly charmed tetraquark}},\ }\href {https://doi.org/10.1038/s41567-022-01614-y} {\bibfield  {journal} {\bibinfo  {journal} {Nature Phys.}\ }\textbf {\bibinfo {volume} {18}},\ \bibinfo {pages} {751} (\bibinfo {year} {2022}{\natexlab{a}})},\ \Eprint {https://arxiv.org/abs/2109.01038} {arXiv:2109.01038 [hep-ex]} \BibitemShut {NoStop}%
\bibitem [{\citenamefont {Aaij}\ \emph {et~al.}(2022{\natexlab{b}})\citenamefont {Aaij} \emph {et~al.}}]{LHCb:2021auc}%
  \BibitemOpen
  \bibfield  {author} {\bibinfo {author} {\bibfnamefont {R.}~\bibnamefont {Aaij}} \emph {et~al.} (\bibinfo {collaboration} {LHCb}),\ }\bibfield  {title} {\bibinfo {title} {{Study of the doubly charmed tetraquark $T_{cc}^{+}$}},\ }\href {https://doi.org/10.1038/s41467-022-30206-w} {\bibfield  {journal} {\bibinfo  {journal} {Nature Commun.}\ }\textbf {\bibinfo {volume} {13}},\ \bibinfo {pages} {3351} (\bibinfo {year} {2022}{\natexlab{b}})},\ \Eprint {https://arxiv.org/abs/2109.01056} {arXiv:2109.01056 [hep-ex]} \BibitemShut {NoStop}%
\bibitem [{\citenamefont {Albaladejo}(2021)}]{Albaladejo:2021vln}%
  \BibitemOpen
  \bibfield  {author} {\bibinfo {author} {\bibfnamefont {M.}~\bibnamefont {Albaladejo}},\ }\bibfield  {title} {\bibinfo {title} {{$T_{cc}^{+}$ coupled channel analysis and predictions}},\ }\href@noop {} {\  (\bibinfo {year} {2021})},\ \Eprint {https://arxiv.org/abs/2110.02944} {arXiv:2110.02944 [hep-ph]} \BibitemShut {NoStop}%
\bibitem [{\citenamefont {Meng}\ \emph {et~al.}(2021)\citenamefont {Meng}, \citenamefont {Wang}, \citenamefont {Wang},\ and\ \citenamefont {Zhu}}]{Meng:2021jnw}%
  \BibitemOpen
  \bibfield  {author} {\bibinfo {author} {\bibfnamefont {L.}~\bibnamefont {Meng}}, \bibinfo {author} {\bibfnamefont {G.-J.}\ \bibnamefont {Wang}}, \bibinfo {author} {\bibfnamefont {B.}~\bibnamefont {Wang}},\ and\ \bibinfo {author} {\bibfnamefont {S.-L.}\ \bibnamefont {Zhu}},\ }\bibfield  {title} {\bibinfo {title} {{Probing the long-range structure of the Tcc+ with the strong and electromagnetic decays}},\ }\href {https://doi.org/10.1103/PhysRevD.104.L051502} {\bibfield  {journal} {\bibinfo  {journal} {Phys. Rev. D}\ }\textbf {\bibinfo {volume} {104}},\ \bibinfo {pages} {051502} (\bibinfo {year} {2021})},\ \Eprint {https://arxiv.org/abs/2107.14784} {arXiv:2107.14784 [hep-ph]} \BibitemShut {NoStop}%
\bibitem [{\citenamefont {Du}\ \emph {et~al.}(2022)\citenamefont {Du}, \citenamefont {Baru}, \citenamefont {Dong}, \citenamefont {Filin}, \citenamefont {Guo}, \citenamefont {Hanhart}, \citenamefont {Nefediev}, \citenamefont {Nieves},\ and\ \citenamefont {Wang}}]{Du:2021zzh}%
  \BibitemOpen
  \bibfield  {author} {\bibinfo {author} {\bibfnamefont {M.-L.}\ \bibnamefont {Du}}, \bibinfo {author} {\bibfnamefont {V.}~\bibnamefont {Baru}}, \bibinfo {author} {\bibfnamefont {X.-K.}\ \bibnamefont {Dong}}, \bibinfo {author} {\bibfnamefont {A.}~\bibnamefont {Filin}}, \bibinfo {author} {\bibfnamefont {F.-K.}\ \bibnamefont {Guo}}, \bibinfo {author} {\bibfnamefont {C.}~\bibnamefont {Hanhart}}, \bibinfo {author} {\bibfnamefont {A.}~\bibnamefont {Nefediev}}, \bibinfo {author} {\bibfnamefont {J.}~\bibnamefont {Nieves}},\ and\ \bibinfo {author} {\bibfnamefont {Q.}~\bibnamefont {Wang}},\ }\bibfield  {title} {\bibinfo {title} {{Coupled-channel approach to $T_{cc}^+$ including three-body effects}},\ }\href {https://doi.org/10.1103/PhysRevD.105.014024} {\bibfield  {journal} {\bibinfo  {journal} {Phys. Rev. D}\ }\textbf {\bibinfo {volume} {105}},\ \bibinfo {pages} {014024} (\bibinfo {year} {2022})},\ \Eprint {https://arxiv.org/abs/2110.13765} {arXiv:2110.13765 [hep-ph]} \BibitemShut {NoStop}%
\bibitem [{\citenamefont {Braaten}\ \emph {et~al.}(2022)\citenamefont {Braaten}, \citenamefont {He}, \citenamefont {Ingles},\ and\ \citenamefont {Jiang}}]{Braaten:2022elw}%
  \BibitemOpen
  \bibfield  {author} {\bibinfo {author} {\bibfnamefont {E.}~\bibnamefont {Braaten}}, \bibinfo {author} {\bibfnamefont {L.-P.}\ \bibnamefont {He}}, \bibinfo {author} {\bibfnamefont {K.}~\bibnamefont {Ingles}},\ and\ \bibinfo {author} {\bibfnamefont {J.}~\bibnamefont {Jiang}},\ }\bibfield  {title} {\bibinfo {title} {{Triangle singularity in the production of Tcc+(3875) and a soft pion}},\ }\href {https://doi.org/10.1103/PhysRevD.106.034033} {\bibfield  {journal} {\bibinfo  {journal} {Phys. Rev. D}\ }\textbf {\bibinfo {volume} {106}},\ \bibinfo {pages} {034033} (\bibinfo {year} {2022})},\ \Eprint {https://arxiv.org/abs/2202.03900} {arXiv:2202.03900 [hep-ph]} \BibitemShut {NoStop}%
\bibitem [{\citenamefont {Wang}\ and\ \citenamefont {Meng}(2023)}]{Wang:2022jop}%
  \BibitemOpen
  \bibfield  {author} {\bibinfo {author} {\bibfnamefont {B.}~\bibnamefont {Wang}}\ and\ \bibinfo {author} {\bibfnamefont {L.}~\bibnamefont {Meng}},\ }\bibfield  {title} {\bibinfo {title} {{Revisiting the DD* chiral interactions with the local momentum-space regularization up to the third order and the nature of Tcc+}},\ }\href {https://doi.org/10.1103/PhysRevD.107.094002} {\bibfield  {journal} {\bibinfo  {journal} {Phys. Rev. D}\ }\textbf {\bibinfo {volume} {107}},\ \bibinfo {pages} {094002} (\bibinfo {year} {2023})},\ \Eprint {https://arxiv.org/abs/2212.08447} {arXiv:2212.08447 [hep-ph]} \BibitemShut {NoStop}%
\bibitem [{\citenamefont {Dai}\ \emph {et~al.}(2023)\citenamefont {Dai}, \citenamefont {Fleming}, \citenamefont {Hodges},\ and\ \citenamefont {Mehen}}]{Dai:2023mxm}%
  \BibitemOpen
  \bibfield  {author} {\bibinfo {author} {\bibfnamefont {L.}~\bibnamefont {Dai}}, \bibinfo {author} {\bibfnamefont {S.}~\bibnamefont {Fleming}}, \bibinfo {author} {\bibfnamefont {R.}~\bibnamefont {Hodges}},\ and\ \bibinfo {author} {\bibfnamefont {T.}~\bibnamefont {Mehen}},\ }\bibfield  {title} {\bibinfo {title} {{Strong decays of Tcc+ at NLO in an effective field theory}},\ }\href {https://doi.org/10.1103/PhysRevD.107.076001} {\bibfield  {journal} {\bibinfo  {journal} {Phys. Rev. D}\ }\textbf {\bibinfo {volume} {107}},\ \bibinfo {pages} {076001} (\bibinfo {year} {2023})},\ \Eprint {https://arxiv.org/abs/2301.11950} {arXiv:2301.11950 [hep-ph]} \BibitemShut {NoStop}%
\bibitem [{\citenamefont {Padmanath}\ and\ \citenamefont {Prelovsek}(2022)}]{Padmanath:2022cvl}%
  \BibitemOpen
  \bibfield  {author} {\bibinfo {author} {\bibfnamefont {M.}~\bibnamefont {Padmanath}}\ and\ \bibinfo {author} {\bibfnamefont {S.}~\bibnamefont {Prelovsek}},\ }\bibfield  {title} {\bibinfo {title} {{Signature of a Doubly Charm Tetraquark Pole in $DD^*$ Scattering on the Lattice}},\ }\href {https://doi.org/10.1103/PhysRevLett.129.032002} {\bibfield  {journal} {\bibinfo  {journal} {Phys. Rev. Lett.}\ }\textbf {\bibinfo {volume} {129}},\ \bibinfo {pages} {032002} (\bibinfo {year} {2022})},\ \Eprint {https://arxiv.org/abs/2202.10110} {arXiv:2202.10110 [hep-lat]} \BibitemShut {NoStop}%
\bibitem [{\citenamefont {Chen}\ \emph {et~al.}(2022)\citenamefont {Chen}, \citenamefont {Shi}, \citenamefont {Chen}, \citenamefont {Gong}, \citenamefont {Liu}, \citenamefont {Sun},\ and\ \citenamefont {Zhang}}]{Chen:2022vpo}%
  \BibitemOpen
  \bibfield  {author} {\bibinfo {author} {\bibfnamefont {S.}~\bibnamefont {Chen}}, \bibinfo {author} {\bibfnamefont {C.}~\bibnamefont {Shi}}, \bibinfo {author} {\bibfnamefont {Y.}~\bibnamefont {Chen}}, \bibinfo {author} {\bibfnamefont {M.}~\bibnamefont {Gong}}, \bibinfo {author} {\bibfnamefont {Z.}~\bibnamefont {Liu}}, \bibinfo {author} {\bibfnamefont {W.}~\bibnamefont {Sun}},\ and\ \bibinfo {author} {\bibfnamefont {R.}~\bibnamefont {Zhang}},\ }\bibfield  {title} {\bibinfo {title} {{$T_{cc}^+(3875)$ relevant $DD^*$ scattering from $N_f=2$ lattice QCD}},\ }\href {https://doi.org/10.1016/j.physletb.2022.137391} {\bibfield  {journal} {\bibinfo  {journal} {Phys. Lett. B}\ }\textbf {\bibinfo {volume} {833}},\ \bibinfo {pages} {137391} (\bibinfo {year} {2022})},\ \Eprint {https://arxiv.org/abs/2206.06185} {arXiv:2206.06185 [hep-lat]} \BibitemShut {NoStop}%
\bibitem [{\citenamefont {Lyu}\ \emph {et~al.}(2023)\citenamefont {Lyu}, \citenamefont {Aoki}, \citenamefont {Doi}, \citenamefont {Hatsuda}, \citenamefont {Ikeda},\ and\ \citenamefont {Meng}}]{Lyu:2023xro}%
  \BibitemOpen
  \bibfield  {author} {\bibinfo {author} {\bibfnamefont {Y.}~\bibnamefont {Lyu}}, \bibinfo {author} {\bibfnamefont {S.}~\bibnamefont {Aoki}}, \bibinfo {author} {\bibfnamefont {T.}~\bibnamefont {Doi}}, \bibinfo {author} {\bibfnamefont {T.}~\bibnamefont {Hatsuda}}, \bibinfo {author} {\bibfnamefont {Y.}~\bibnamefont {Ikeda}},\ and\ \bibinfo {author} {\bibfnamefont {J.}~\bibnamefont {Meng}},\ }\bibfield  {title} {\bibinfo {title} {{Doubly Charmed Tetraquark Tcc+ from Lattice QCD near Physical Point}},\ }\href {https://doi.org/10.1103/PhysRevLett.131.161901} {\bibfield  {journal} {\bibinfo  {journal} {Phys. Rev. Lett.}\ }\textbf {\bibinfo {volume} {131}},\ \bibinfo {pages} {161901} (\bibinfo {year} {2023})},\ \Eprint {https://arxiv.org/abs/2302.04505} {arXiv:2302.04505 [hep-lat]} \BibitemShut {NoStop}%
\bibitem [{\citenamefont {Matuschek}\ \emph {et~al.}(2021)\citenamefont {Matuschek}, \citenamefont {Baru}, \citenamefont {Guo},\ and\ \citenamefont {Hanhart}}]{Matuschek:2020gqe}%
  \BibitemOpen
  \bibfield  {author} {\bibinfo {author} {\bibfnamefont {I.}~\bibnamefont {Matuschek}}, \bibinfo {author} {\bibfnamefont {V.}~\bibnamefont {Baru}}, \bibinfo {author} {\bibfnamefont {F.-K.}\ \bibnamefont {Guo}},\ and\ \bibinfo {author} {\bibfnamefont {C.}~\bibnamefont {Hanhart}},\ }\bibfield  {title} {\bibinfo {title} {{On the nature of near-threshold bound and virtual states}},\ }\href {https://doi.org/10.1140/epja/s10050-021-00413-y} {\bibfield  {journal} {\bibinfo  {journal} {Eur. Phys. J. A}\ }\textbf {\bibinfo {volume} {57}},\ \bibinfo {pages} {101} (\bibinfo {year} {2021})},\ \Eprint {https://arxiv.org/abs/2007.05329} {arXiv:2007.05329 [hep-ph]} \BibitemShut {NoStop}%
\bibitem [{\citenamefont {Du}\ \emph {et~al.}(2023)\citenamefont {Du}, \citenamefont {Filin}, \citenamefont {Baru}, \citenamefont {Dong}, \citenamefont {Epelbaum}, \citenamefont {Guo}, \citenamefont {Hanhart}, \citenamefont {Nefediev}, \citenamefont {Nieves},\ and\ \citenamefont {Wang}}]{Du:2023hlu}%
  \BibitemOpen
  \bibfield  {author} {\bibinfo {author} {\bibfnamefont {M.-L.}\ \bibnamefont {Du}}, \bibinfo {author} {\bibfnamefont {A.}~\bibnamefont {Filin}}, \bibinfo {author} {\bibfnamefont {V.}~\bibnamefont {Baru}}, \bibinfo {author} {\bibfnamefont {X.-K.}\ \bibnamefont {Dong}}, \bibinfo {author} {\bibfnamefont {E.}~\bibnamefont {Epelbaum}}, \bibinfo {author} {\bibfnamefont {F.-K.}\ \bibnamefont {Guo}}, \bibinfo {author} {\bibfnamefont {C.}~\bibnamefont {Hanhart}}, \bibinfo {author} {\bibfnamefont {A.}~\bibnamefont {Nefediev}}, \bibinfo {author} {\bibfnamefont {J.}~\bibnamefont {Nieves}},\ and\ \bibinfo {author} {\bibfnamefont {Q.}~\bibnamefont {Wang}},\ }\bibfield  {title} {\bibinfo {title} {{Role of Left-Hand Cut Contributions on Pole Extractions from Lattice Data: Case Study for Tcc(3875)+}},\ }\href {https://doi.org/10.1103/PhysRevLett.131.131903} {\bibfield  {journal} {\bibinfo  {journal} {Phys. Rev. Lett.}\ }\textbf {\bibinfo {volume} {131}},\ \bibinfo {pages} {131903} (\bibinfo {year} {2023})},\ \Eprint
  {https://arxiv.org/abs/2303.09441} {arXiv:2303.09441 [hep-ph]} \BibitemShut {NoStop}%
\bibitem [{\citenamefont {Luscher}(1986)}]{Luscher:1986pf}%
  \BibitemOpen
  \bibfield  {author} {\bibinfo {author} {\bibfnamefont {M.}~\bibnamefont {Luscher}},\ }\bibfield  {title} {\bibinfo {title} {{Volume Dependence of the Energy Spectrum in Massive Quantum Field Theories. 2. Scattering States}},\ }\href {https://doi.org/10.1007/BF01211097} {\bibfield  {journal} {\bibinfo  {journal} {Commun. Math. Phys.}\ }\textbf {\bibinfo {volume} {105}},\ \bibinfo {pages} {153} (\bibinfo {year} {1986})}\BibitemShut {NoStop}%
\bibitem [{\citenamefont {Luscher}(1991)}]{Luscher:1990ux}%
  \BibitemOpen
  \bibfield  {author} {\bibinfo {author} {\bibfnamefont {M.}~\bibnamefont {Luscher}},\ }\bibfield  {title} {\bibinfo {title} {{Two particle states on a torus and their relation to the scattering matrix}},\ }\href {https://doi.org/10.1016/0550-3213(91)90366-6} {\bibfield  {journal} {\bibinfo  {journal} {Nucl. Phys. B}\ }\textbf {\bibinfo {volume} {354}},\ \bibinfo {pages} {531} (\bibinfo {year} {1991})}\BibitemShut {NoStop}%
\bibitem [{\citenamefont {Kim}\ \emph {et~al.}(2005)\citenamefont {Kim}, \citenamefont {Sachrajda},\ and\ \citenamefont {Sharpe}}]{Kim:2005gf}%
  \BibitemOpen
  \bibfield  {author} {\bibinfo {author} {\bibfnamefont {C.~H.}\ \bibnamefont {Kim}}, \bibinfo {author} {\bibfnamefont {C.~T.}\ \bibnamefont {Sachrajda}},\ and\ \bibinfo {author} {\bibfnamefont {S.~R.}\ \bibnamefont {Sharpe}},\ }\bibfield  {title} {\bibinfo {title} {{Finite-volume effects for two-hadron states in moving frames}},\ }\href {https://doi.org/10.1016/j.nuclphysb.2005.08.029} {\bibfield  {journal} {\bibinfo  {journal} {Nucl. Phys. B}\ }\textbf {\bibinfo {volume} {727}},\ \bibinfo {pages} {218} (\bibinfo {year} {2005})},\ \Eprint {https://arxiv.org/abs/hep-lat/0507006} {arXiv:hep-lat/0507006} \BibitemShut {NoStop}%
\bibitem [{\citenamefont {Briceno}(2014)}]{Briceno:2014oea}%
  \BibitemOpen
  \bibfield  {author} {\bibinfo {author} {\bibfnamefont {R.~A.}\ \bibnamefont {Briceno}},\ }\bibfield  {title} {\bibinfo {title} {{Two-particle multichannel systems in a finite volume with arbitrary spin}},\ }\href {https://doi.org/10.1103/PhysRevD.89.074507} {\bibfield  {journal} {\bibinfo  {journal} {Phys. Rev. D}\ }\textbf {\bibinfo {volume} {89}},\ \bibinfo {pages} {074507} (\bibinfo {year} {2014})},\ \Eprint {https://arxiv.org/abs/1401.3312} {arXiv:1401.3312 [hep-lat]} \BibitemShut {NoStop}%
\bibitem [{\citenamefont {Raposo}\ and\ \citenamefont {Hansen}(2023)}]{Raposo:2023oru}%
  \BibitemOpen
  \bibfield  {author} {\bibinfo {author} {\bibfnamefont {A.~B.~a.}\ \bibnamefont {Raposo}}\ and\ \bibinfo {author} {\bibfnamefont {M.~T.}\ \bibnamefont {Hansen}},\ }\bibfield  {title} {\bibinfo {title} {{Finite-volume scattering on the left-hand cut}},\ }\href@noop {} {\  (\bibinfo {year} {2023})},\ \Eprint {https://arxiv.org/abs/2311.18793} {arXiv:2311.18793 [hep-lat]} \BibitemShut {NoStop}%
\bibitem [{\citenamefont {Dawid}\ \emph {et~al.}(2023)\citenamefont {Dawid}, \citenamefont {Islam},\ and\ \citenamefont {Brice\~no}}]{Dawid:2023jrj}%
  \BibitemOpen
  \bibfield  {author} {\bibinfo {author} {\bibfnamefont {S.~M.}\ \bibnamefont {Dawid}}, \bibinfo {author} {\bibfnamefont {M.~H.~E.}\ \bibnamefont {Islam}},\ and\ \bibinfo {author} {\bibfnamefont {R.~A.}\ \bibnamefont {Brice\~no}},\ }\bibfield  {title} {\bibinfo {title} {{Analytic continuation of the relativistic three-particle scattering amplitudes}},\ }\href@noop {} {\  (\bibinfo {year} {2023})},\ \Eprint {https://arxiv.org/abs/2303.04394} {arXiv:2303.04394 [nucl-th]} \BibitemShut {NoStop}%
\bibitem [{\citenamefont {Green}\ \emph {et~al.}(2021)\citenamefont {Green}, \citenamefont {Hanlon}, \citenamefont {Junnarkar},\ and\ \citenamefont {Wittig}}]{Green:2021qol}%
  \BibitemOpen
  \bibfield  {author} {\bibinfo {author} {\bibfnamefont {J.~R.}\ \bibnamefont {Green}}, \bibinfo {author} {\bibfnamefont {A.~D.}\ \bibnamefont {Hanlon}}, \bibinfo {author} {\bibfnamefont {P.~M.}\ \bibnamefont {Junnarkar}},\ and\ \bibinfo {author} {\bibfnamefont {H.}~\bibnamefont {Wittig}},\ }\bibfield  {title} {\bibinfo {title} {{Weakly bound $H$ dibaryon from SU(3)-flavor-symmetric QCD}},\ }\href {https://doi.org/10.1103/PhysRevLett.127.242003} {\bibfield  {journal} {\bibinfo  {journal} {Phys. Rev. Lett.}\ }\textbf {\bibinfo {volume} {127}},\ \bibinfo {pages} {242003} (\bibinfo {year} {2021})},\ \Eprint {https://arxiv.org/abs/2103.01054} {arXiv:2103.01054 [hep-lat]} \BibitemShut {NoStop}%
\bibitem [{\citenamefont {Sato}\ and\ \citenamefont {Bedaque}(2007)}]{Sato:2007ms}%
  \BibitemOpen
  \bibfield  {author} {\bibinfo {author} {\bibfnamefont {I.}~\bibnamefont {Sato}}\ and\ \bibinfo {author} {\bibfnamefont {P.~F.}\ \bibnamefont {Bedaque}},\ }\bibfield  {title} {\bibinfo {title} {{Fitting two nucleons inside a box: Exponentially suppressed corrections to the Luscher's formula}},\ }\href {https://doi.org/10.1103/PhysRevD.76.034502} {\bibfield  {journal} {\bibinfo  {journal} {Phys. Rev. D}\ }\textbf {\bibinfo {volume} {76}},\ \bibinfo {pages} {034502} (\bibinfo {year} {2007})},\ \Eprint {https://arxiv.org/abs/hep-lat/0702021} {arXiv:hep-lat/0702021} \BibitemShut {NoStop}%
\bibitem [{\citenamefont {Meng}\ and\ \citenamefont {Epelbaum}(2021)}]{Meng:2021uhz}%
  \BibitemOpen
  \bibfield  {author} {\bibinfo {author} {\bibfnamefont {L.}~\bibnamefont {Meng}}\ and\ \bibinfo {author} {\bibfnamefont {E.}~\bibnamefont {Epelbaum}},\ }\bibfield  {title} {\bibinfo {title} {{Two-particle scattering from finite-volume quantization conditions using the plane wave basis}},\ }\href {https://doi.org/10.1007/JHEP10(2021)051} {\bibfield  {journal} {\bibinfo  {journal} {JHEP}\ }\textbf {\bibinfo {volume} {10}},\ \bibinfo {pages} {051}},\ \Eprint {https://arxiv.org/abs/2108.02709} {arXiv:2108.02709 [hep-lat]} \BibitemShut {NoStop}%
\bibitem [{\citenamefont {Leskovec}\ and\ \citenamefont {Prelovsek}(2012)}]{Leskovec:2012gb}%
  \BibitemOpen
  \bibfield  {author} {\bibinfo {author} {\bibfnamefont {L.}~\bibnamefont {Leskovec}}\ and\ \bibinfo {author} {\bibfnamefont {S.}~\bibnamefont {Prelovsek}},\ }\bibfield  {title} {\bibinfo {title} {{Scattering phase shifts for two particles of different mass and non-zero total momentum in lattice QCD}},\ }\href {https://doi.org/10.1103/PhysRevD.85.114507} {\bibfield  {journal} {\bibinfo  {journal} {Phys. Rev. D}\ }\textbf {\bibinfo {volume} {85}},\ \bibinfo {pages} {114507} (\bibinfo {year} {2012})},\ \Eprint {https://arxiv.org/abs/1202.2145} {arXiv:1202.2145 [hep-lat]} \BibitemShut {NoStop}%
\bibitem [{\citenamefont {Dudek}\ \emph {et~al.}(2012)\citenamefont {Dudek}, \citenamefont {Edwards},\ and\ \citenamefont {Thomas}}]{Dudek:2012gj}%
  \BibitemOpen
  \bibfield  {author} {\bibinfo {author} {\bibfnamefont {J.~J.}\ \bibnamefont {Dudek}}, \bibinfo {author} {\bibfnamefont {R.~G.}\ \bibnamefont {Edwards}},\ and\ \bibinfo {author} {\bibfnamefont {C.~E.}\ \bibnamefont {Thomas}},\ }\bibfield  {title} {\bibinfo {title} {{S and D-wave phase shifts in isospin-2 pi pi scattering from lattice QCD}},\ }\href {https://doi.org/10.1103/PhysRevD.86.034031} {\bibfield  {journal} {\bibinfo  {journal} {Phys. Rev. D}\ }\textbf {\bibinfo {volume} {86}},\ \bibinfo {pages} {034031} (\bibinfo {year} {2012})},\ \Eprint {https://arxiv.org/abs/1203.6041} {arXiv:1203.6041 [hep-ph]} \BibitemShut {NoStop}%
\bibitem [{\citenamefont {Woss}\ \emph {et~al.}(2018)\citenamefont {Woss}, \citenamefont {Thomas}, \citenamefont {Dudek}, \citenamefont {Edwards},\ and\ \citenamefont {Wilson}}]{Woss:2018irj}%
  \BibitemOpen
  \bibfield  {author} {\bibinfo {author} {\bibfnamefont {A.}~\bibnamefont {Woss}}, \bibinfo {author} {\bibfnamefont {C.~E.}\ \bibnamefont {Thomas}}, \bibinfo {author} {\bibfnamefont {J.~J.}\ \bibnamefont {Dudek}}, \bibinfo {author} {\bibfnamefont {R.~G.}\ \bibnamefont {Edwards}},\ and\ \bibinfo {author} {\bibfnamefont {D.~J.}\ \bibnamefont {Wilson}},\ }\bibfield  {title} {\bibinfo {title} {{Dynamically-coupled partial-waves in $\rho\pi$ isospin-2 scattering from lattice QCD}},\ }\href {https://doi.org/10.1007/JHEP07(2018)043} {\bibfield  {journal} {\bibinfo  {journal} {JHEP}\ }\textbf {\bibinfo {volume} {07}},\ \bibinfo {pages} {043}},\ \Eprint {https://arxiv.org/abs/1802.05580} {arXiv:1802.05580 [hep-lat]} \BibitemShut {NoStop}%
\bibitem [{\citenamefont {Morningstar}\ \emph {et~al.}(2017)\citenamefont {Morningstar}, \citenamefont {Bulava}, \citenamefont {Singha}, \citenamefont {Brett}, \citenamefont {Fallica}, \citenamefont {Hanlon},\ and\ \citenamefont {H\"orz}}]{Morningstar:2017spu}%
  \BibitemOpen
  \bibfield  {author} {\bibinfo {author} {\bibfnamefont {C.}~\bibnamefont {Morningstar}}, \bibinfo {author} {\bibfnamefont {J.}~\bibnamefont {Bulava}}, \bibinfo {author} {\bibfnamefont {B.}~\bibnamefont {Singha}}, \bibinfo {author} {\bibfnamefont {R.}~\bibnamefont {Brett}}, \bibinfo {author} {\bibfnamefont {J.}~\bibnamefont {Fallica}}, \bibinfo {author} {\bibfnamefont {A.}~\bibnamefont {Hanlon}},\ and\ \bibinfo {author} {\bibfnamefont {B.}~\bibnamefont {H\"orz}},\ }\bibfield  {title} {\bibinfo {title} {{Estimating the two-particle $K$-matrix for multiple partial waves and decay channels from finite-volume energies}},\ }\href {https://doi.org/10.1016/j.nuclphysb.2017.09.014} {\bibfield  {journal} {\bibinfo  {journal} {Nucl. Phys. B}\ }\textbf {\bibinfo {volume} {924}},\ \bibinfo {pages} {477} (\bibinfo {year} {2017})},\ \Eprint {https://arxiv.org/abs/1707.05817} {arXiv:1707.05817 [hep-lat]} \BibitemShut {NoStop}%
\bibitem [{\citenamefont {Meng}\ \emph {et~al.}(2023{\natexlab{b}})\citenamefont {Meng}, \citenamefont {Baru}, \citenamefont {Epelbaum}, \citenamefont {Filin},\ and\ \citenamefont {Gasparyan}}]{Suppl_Tcc_EFV}%
  \BibitemOpen
  \bibfield  {author} {\bibinfo {author} {\bibfnamefont {L.}~\bibnamefont {Meng}}, \bibinfo {author} {\bibfnamefont {V.}~\bibnamefont {Baru}}, \bibinfo {author} {\bibfnamefont {E.}~\bibnamefont {Epelbaum}}, \bibinfo {author} {\bibfnamefont {A.}~\bibnamefont {Filin}},\ and\ \bibinfo {author} {\bibfnamefont {A.}~\bibnamefont {Gasparyan}},\ }\bibfield  {title} {\bibinfo {title} {{See Supplemental Material for additional details on our framework, the strategy employed to determine the $D^*D\pi$ coupling constant, fitting procedure and the estimation of systematic uncertainties.}},\ }\href@noop {} {\  (\bibinfo {year} {2023}{\natexlab{b}})}\BibitemShut {NoStop}%
\bibitem [{\citenamefont {Becirevic}\ and\ \citenamefont {Sanfilippo}(2013)}]{Becirevic:2012pf}%
  \BibitemOpen
  \bibfield  {author} {\bibinfo {author} {\bibfnamefont {D.}~\bibnamefont {Becirevic}}\ and\ \bibinfo {author} {\bibfnamefont {F.}~\bibnamefont {Sanfilippo}},\ }\bibfield  {title} {\bibinfo {title} {{Theoretical estimate of the $D^* \to D\pi$ decay rate}},\ }\href {https://doi.org/10.1016/j.physletb.2013.03.004} {\bibfield  {journal} {\bibinfo  {journal} {Phys. Lett. B}\ }\textbf {\bibinfo {volume} {721}},\ \bibinfo {pages} {94} (\bibinfo {year} {2013})},\ \Eprint {https://arxiv.org/abs/1210.5410} {arXiv:1210.5410 [hep-lat]} \BibitemShut {NoStop}%
\bibitem [{\citenamefont {Reinert}\ \emph {et~al.}(2018)\citenamefont {Reinert}, \citenamefont {Krebs},\ and\ \citenamefont {Epelbaum}}]{Reinert:2017usi}%
  \BibitemOpen
  \bibfield  {author} {\bibinfo {author} {\bibfnamefont {P.}~\bibnamefont {Reinert}}, \bibinfo {author} {\bibfnamefont {H.}~\bibnamefont {Krebs}},\ and\ \bibinfo {author} {\bibfnamefont {E.}~\bibnamefont {Epelbaum}},\ }\bibfield  {title} {\bibinfo {title} {{Semilocal momentum-space regularized chiral two-nucleon potentials up to fifth order}},\ }\href {https://doi.org/10.1140/epja/i2018-12516-4} {\bibfield  {journal} {\bibinfo  {journal} {Eur. Phys. J. A}\ }\textbf {\bibinfo {volume} {54}},\ \bibinfo {pages} {86} (\bibinfo {year} {2018})},\ \Eprint {https://arxiv.org/abs/1711.08821} {arXiv:1711.08821 [nucl-th]} \BibitemShut {NoStop}%
\bibitem [{\citenamefont {Hansen}\ \emph {et~al.}(2024)\citenamefont {Hansen}, \citenamefont {Romero-L\'opez},\ and\ \citenamefont {Sharpe}}]{Hansen:2024ffk}%
  \BibitemOpen
  \bibfield  {author} {\bibinfo {author} {\bibfnamefont {M.~T.}\ \bibnamefont {Hansen}}, \bibinfo {author} {\bibfnamefont {F.}~\bibnamefont {Romero-L\'opez}},\ and\ \bibinfo {author} {\bibfnamefont {S.~R.}\ \bibnamefont {Sharpe}},\ }\bibfield  {title} {\bibinfo {title} {{Incorporating $DD\pi$ effects and left-hand cuts in lattice QCD studies of the $T_{cc}(3875)^+$}},\ }\href@noop {} {\  (\bibinfo {year} {2024})},\ \Eprint {https://arxiv.org/abs/2401.06609} {arXiv:2401.06609 [hep-lat]} \BibitemShut {NoStop}%
\bibitem [{\citenamefont {Prelovsek}\ and\ \citenamefont {Leskovec}(2013)}]{Prelovsek:2013cra}%
  \BibitemOpen
  \bibfield  {author} {\bibinfo {author} {\bibfnamefont {S.}~\bibnamefont {Prelovsek}}\ and\ \bibinfo {author} {\bibfnamefont {L.}~\bibnamefont {Leskovec}},\ }\bibfield  {title} {\bibinfo {title} {{Evidence for $X(3872)$ from $DD^*$ scattering on the lattice}},\ }\href {https://doi.org/10.1103/PhysRevLett.111.192001} {\bibfield  {journal} {\bibinfo  {journal} {Phys. Rev. Lett.}\ }\textbf {\bibinfo {volume} {111}},\ \bibinfo {pages} {192001} (\bibinfo {year} {2013})},\ \Eprint {https://arxiv.org/abs/1307.5172} {arXiv:1307.5172 [hep-lat]} \BibitemShut {NoStop}%
\bibitem [{\citenamefont {Kadyshevsky}(1968)}]{Kadyshevsky:1967rs}%
  \BibitemOpen
  \bibfield  {author} {\bibinfo {author} {\bibfnamefont {V.~G.}\ \bibnamefont {Kadyshevsky}},\ }\bibfield  {title} {\bibinfo {title} {{Quasipotential type equation for the relativistic scattering amplitude}},\ }\href {https://doi.org/10.1016/0550-3213(68)90274-5} {\bibfield  {journal} {\bibinfo  {journal} {Nucl. Phys. B}\ }\textbf {\bibinfo {volume} {6}},\ \bibinfo {pages} {125} (\bibinfo {year} {1968})}\BibitemShut {NoStop}%
\bibitem [{\citenamefont {Baru}\ \emph {et~al.}(2019{\natexlab{a}})\citenamefont {Baru}, \citenamefont {Epelbaum}, \citenamefont {Gegelia},\ and\ \citenamefont {Ren}}]{Baru:2019ndr}%
  \BibitemOpen
  \bibfield  {author} {\bibinfo {author} {\bibfnamefont {V.}~\bibnamefont {Baru}}, \bibinfo {author} {\bibfnamefont {E.}~\bibnamefont {Epelbaum}}, \bibinfo {author} {\bibfnamefont {J.}~\bibnamefont {Gegelia}},\ and\ \bibinfo {author} {\bibfnamefont {X.~L.}\ \bibnamefont {Ren}},\ }\bibfield  {title} {\bibinfo {title} {{Towards baryon-baryon scattering in manifestly Lorentz-invariant formulation of SU(3) baryon chiral perturbation theory}},\ }\href {https://doi.org/10.1016/j.physletb.2019.134987} {\bibfield  {journal} {\bibinfo  {journal} {Phys. Lett. B}\ }\textbf {\bibinfo {volume} {798}},\ \bibinfo {pages} {134987} (\bibinfo {year} {2019}{\natexlab{a}})},\ \Eprint {https://arxiv.org/abs/1905.02116} {arXiv:1905.02116 [nucl-th]} \BibitemShut {NoStop}%
\bibitem [{\citenamefont {Woloshyn}\ and\ \citenamefont {Jackson}(1973)}]{Woloshyn:1973mce}%
  \BibitemOpen
  \bibfield  {author} {\bibinfo {author} {\bibfnamefont {R.~M.}\ \bibnamefont {Woloshyn}}\ and\ \bibinfo {author} {\bibfnamefont {A.~D.}\ \bibnamefont {Jackson}},\ }\bibfield  {title} {\bibinfo {title} {{Comparison of three-dimensional relativistic scattering equations}},\ }\href {https://doi.org/10.1016/0550-3213(73)90626-3} {\bibfield  {journal} {\bibinfo  {journal} {Nucl. Phys. B}\ }\textbf {\bibinfo {volume} {64}},\ \bibinfo {pages} {269} (\bibinfo {year} {1973})}\BibitemShut {NoStop}%
\bibitem [{Note1()}]{Note1}%
  \BibitemOpen
  \bibinfo {note} {Neglecting the last term in the second line in Eq.~\protect \eqref {eq:sm_G_sup} corresponds to the Kadyshevsky equation \cite {Kadyshevsky:1967rs} (see also \cite {Baru:2019ndr} for a related discussion). We checked that omitting this term does not have an impact on our results.}\BibitemShut {Stop}%
\bibitem [{\citenamefont {Zhang}\ \emph {et~al.}(2022)\citenamefont {Zhang}, \citenamefont {Hanhart}, \citenamefont {Mei\ss{}ner},\ and\ \citenamefont {Xie}}]{Zhang:2021hcl}%
  \BibitemOpen
  \bibfield  {author} {\bibinfo {author} {\bibfnamefont {X.}~\bibnamefont {Zhang}}, \bibinfo {author} {\bibfnamefont {C.}~\bibnamefont {Hanhart}}, \bibinfo {author} {\bibfnamefont {U.-G.}\ \bibnamefont {Mei\ss{}ner}},\ and\ \bibinfo {author} {\bibfnamefont {J.-J.}\ \bibnamefont {Xie}},\ }\bibfield  {title} {\bibinfo {title} {{Remarks on non-perturbative three\textendash{}body dynamics and its application to the $KK{\bar{K}}$ system}},\ }\href {https://doi.org/10.1140/epja/s10050-021-00661-y} {\bibfield  {journal} {\bibinfo  {journal} {Eur. Phys. J. A}\ }\textbf {\bibinfo {volume} {58}},\ \bibinfo {pages} {20} (\bibinfo {year} {2022})},\ \Eprint {https://arxiv.org/abs/2107.03168} {arXiv:2107.03168 [hep-ph]} \BibitemShut {NoStop}%
\bibitem [{\citenamefont {Hanhart}\ \emph {et~al.}(2001)\citenamefont {Hanhart}, \citenamefont {Miller}, \citenamefont {Myhrer}, \citenamefont {Sato},\ and\ \citenamefont {van Kolck}}]{Hanhart:2000wf}%
  \BibitemOpen
  \bibfield  {author} {\bibinfo {author} {\bibfnamefont {C.}~\bibnamefont {Hanhart}}, \bibinfo {author} {\bibfnamefont {G.~A.}\ \bibnamefont {Miller}}, \bibinfo {author} {\bibfnamefont {F.}~\bibnamefont {Myhrer}}, \bibinfo {author} {\bibfnamefont {T.}~\bibnamefont {Sato}},\ and\ \bibinfo {author} {\bibfnamefont {U.}~\bibnamefont {van Kolck}},\ }\bibfield  {title} {\bibinfo {title} {{Toy model for pion production in nucleon-nucleon collisions}},\ }\href {https://doi.org/10.1103/PhysRevC.63.044002} {\bibfield  {journal} {\bibinfo  {journal} {Phys. Rev. C}\ }\textbf {\bibinfo {volume} {63}},\ \bibinfo {pages} {044002} (\bibinfo {year} {2001})},\ \Eprint {https://arxiv.org/abs/nucl-th/0010079} {arXiv:nucl-th/0010079} \BibitemShut {NoStop}%
\bibitem [{\citenamefont {Li}\ \emph {et~al.}(2021)\citenamefont {Li}, \citenamefont {Wu}, \citenamefont {Leinweber},\ and\ \citenamefont {Thomas}}]{Li:2021mob}%
  \BibitemOpen
  \bibfield  {author} {\bibinfo {author} {\bibfnamefont {Y.}~\bibnamefont {Li}}, \bibinfo {author} {\bibfnamefont {J.-J.}\ \bibnamefont {Wu}}, \bibinfo {author} {\bibfnamefont {D.~B.}\ \bibnamefont {Leinweber}},\ and\ \bibinfo {author} {\bibfnamefont {A.~W.}\ \bibnamefont {Thomas}},\ }\bibfield  {title} {\bibinfo {title} {{Hamiltonian effective field theory in elongated or moving finite volume}},\ }\href {https://doi.org/10.1103/PhysRevD.103.094518} {\bibfield  {journal} {\bibinfo  {journal} {Phys. Rev. D}\ }\textbf {\bibinfo {volume} {103}},\ \bibinfo {pages} {094518} (\bibinfo {year} {2021})},\ \Eprint {https://arxiv.org/abs/2103.12260} {arXiv:2103.12260 [hep-lat]} \BibitemShut {NoStop}%
\bibitem [{\citenamefont {Doring}\ \emph {et~al.}(2012)\citenamefont {Doring}, \citenamefont {Mei{\ss}ner}, \citenamefont {Oset},\ and\ \citenamefont {Rusetsky}}]{Doring:2012eu}%
  \BibitemOpen
  \bibfield  {author} {\bibinfo {author} {\bibfnamefont {M.}~\bibnamefont {Doring}}, \bibinfo {author} {\bibfnamefont {U.-G.}\ \bibnamefont {Mei{\ss}ner}}, \bibinfo {author} {\bibfnamefont {E.}~\bibnamefont {Oset}},\ and\ \bibinfo {author} {\bibfnamefont {A.}~\bibnamefont {Rusetsky}},\ }\bibfield  {title} {\bibinfo {title} {{Scalar mesons moving in a finite volume and the role of partial wave mixing}},\ }\href {https://doi.org/10.1140/epja/i2012-12114-6} {\bibfield  {journal} {\bibinfo  {journal} {Eur. Phys. J. A}\ }\textbf {\bibinfo {volume} {48}},\ \bibinfo {pages} {114} (\bibinfo {year} {2012})},\ \Eprint {https://arxiv.org/abs/1205.4838} {arXiv:1205.4838 [hep-lat]} \BibitemShut {NoStop}%
\bibitem [{\citenamefont {Baru}\ \emph {et~al.}(2019{\natexlab{b}})\citenamefont {Baru}, \citenamefont {Epelbaum}, \citenamefont {Filin}, \citenamefont {Hanhart}, \citenamefont {Nefediev},\ and\ \citenamefont {Wang}}]{Baru:2019xnh}%
  \BibitemOpen
  \bibfield  {author} {\bibinfo {author} {\bibfnamefont {V.}~\bibnamefont {Baru}}, \bibinfo {author} {\bibfnamefont {E.}~\bibnamefont {Epelbaum}}, \bibinfo {author} {\bibfnamefont {A.~A.}\ \bibnamefont {Filin}}, \bibinfo {author} {\bibfnamefont {C.}~\bibnamefont {Hanhart}}, \bibinfo {author} {\bibfnamefont {A.~V.}\ \bibnamefont {Nefediev}},\ and\ \bibinfo {author} {\bibfnamefont {Q.}~\bibnamefont {Wang}},\ }\bibfield  {title} {\bibinfo {title} {{Spin partners $W_{bJ}$ from the line shapes of the $Z_b(10610)$ and $Z_b(10650)$}},\ }\href {https://doi.org/10.1103/PhysRevD.99.094013} {\bibfield  {journal} {\bibinfo  {journal} {Phys. Rev. D}\ }\textbf {\bibinfo {volume} {99}},\ \bibinfo {pages} {094013} (\bibinfo {year} {2019}{\natexlab{b}})},\ \Eprint {https://arxiv.org/abs/1901.10319} {arXiv:1901.10319 [hep-ph]} \BibitemShut {NoStop}%
\bibitem [{\citenamefont {Duguet}\ \emph {et~al.}(2023)\citenamefont {Duguet}, \citenamefont {Ekstr\"om}, \citenamefont {Furnstahl}, \citenamefont {K\"onig},\ and\ \citenamefont {Lee}}]{Duguet:2023wuh}%
  \BibitemOpen
  \bibfield  {author} {\bibinfo {author} {\bibfnamefont {T.}~\bibnamefont {Duguet}}, \bibinfo {author} {\bibfnamefont {A.}~\bibnamefont {Ekstr\"om}}, \bibinfo {author} {\bibfnamefont {R.~J.}\ \bibnamefont {Furnstahl}}, \bibinfo {author} {\bibfnamefont {S.}~\bibnamefont {K\"onig}},\ and\ \bibinfo {author} {\bibfnamefont {D.}~\bibnamefont {Lee}},\ }\bibfield  {title} {\bibinfo {title} {{Eigenvector Continuation and Projection-Based Emulators}},\ }\href@noop {} {\  (\bibinfo {year} {2023})},\ \Eprint {https://arxiv.org/abs/2310.19419} {arXiv:2310.19419 [nucl-th]} \BibitemShut {NoStop}%
\bibitem [{\citenamefont {Meng}\ and\ \citenamefont {Epelbaum}(2023)}]{Meng:2023vxy}%
  \BibitemOpen
  \bibfield  {author} {\bibinfo {author} {\bibfnamefont {L.}~\bibnamefont {Meng}}\ and\ \bibinfo {author} {\bibfnamefont {E.}~\bibnamefont {Epelbaum}},\ }\bibfield  {title} {\bibinfo {title} {{Finite volume NN systems using plane wave expansion and eigenvector continuation}},\ }\href {https://doi.org/10.22323/1.430.0201} {\bibfield  {journal} {\bibinfo  {journal} {PoS}\ }\textbf {\bibinfo {volume} {LATTICE2022}},\ \bibinfo {pages} {201} (\bibinfo {year} {2023})}\BibitemShut {NoStop}%
\bibitem [{\citenamefont {Baru}\ \emph {et~al.}(2013)\citenamefont {Baru}, \citenamefont {Epelbaum}, \citenamefont {Filin}, \citenamefont {Hanhart}, \citenamefont {Mei{\ss}ner},\ and\ \citenamefont {Nefediev}}]{Baru:2013rta}%
  \BibitemOpen
  \bibfield  {author} {\bibinfo {author} {\bibfnamefont {V.}~\bibnamefont {Baru}}, \bibinfo {author} {\bibfnamefont {E.}~\bibnamefont {Epelbaum}}, \bibinfo {author} {\bibfnamefont {A.~A.}\ \bibnamefont {Filin}}, \bibinfo {author} {\bibfnamefont {C.}~\bibnamefont {Hanhart}}, \bibinfo {author} {\bibfnamefont {U.-G.}\ \bibnamefont {Mei{\ss}ner}},\ and\ \bibinfo {author} {\bibfnamefont {A.~V.}\ \bibnamefont {Nefediev}},\ }\bibfield  {title} {\bibinfo {title} {{Quark mass dependence of the $X(3872)$ binding energy}},\ }\href {https://doi.org/10.1016/j.physletb.2013.08.073} {\bibfield  {journal} {\bibinfo  {journal} {Phys. Lett. B}\ }\textbf {\bibinfo {volume} {726}},\ \bibinfo {pages} {537} (\bibinfo {year} {2013})},\ \Eprint {https://arxiv.org/abs/1306.4108} {arXiv:1306.4108 [hep-ph]} \BibitemShut {NoStop}%
\bibitem [{\citenamefont {D\"urr}\ \emph {et~al.}(2014)\citenamefont {D\"urr} \emph {et~al.}}]{BMW:2013fzj}%
  \BibitemOpen
  \bibfield  {author} {\bibinfo {author} {\bibfnamefont {S.}~\bibnamefont {D\"urr}} \emph {et~al.} (\bibinfo {collaboration} {BMW}),\ }\bibfield  {title} {\bibinfo {title} {{Lattice QCD at the physical point meets SU(2) chiral perturbation theory}},\ }\href {https://doi.org/10.1103/PhysRevD.90.114504} {\bibfield  {journal} {\bibinfo  {journal} {Phys. Rev. D}\ }\textbf {\bibinfo {volume} {90}},\ \bibinfo {pages} {114504} (\bibinfo {year} {2014})},\ \Eprint {https://arxiv.org/abs/1310.3626} {arXiv:1310.3626 [hep-lat]} \BibitemShut {NoStop}%
\bibitem [{\citenamefont {Casalbuoni}\ \emph {et~al.}(1997)\citenamefont {Casalbuoni}, \citenamefont {Deandrea}, \citenamefont {Di~Bartolomeo}, \citenamefont {Gatto}, \citenamefont {Feruglio},\ and\ \citenamefont {Nardulli}}]{Casalbuoni:1996pg}%
  \BibitemOpen
  \bibfield  {author} {\bibinfo {author} {\bibfnamefont {R.}~\bibnamefont {Casalbuoni}}, \bibinfo {author} {\bibfnamefont {A.}~\bibnamefont {Deandrea}}, \bibinfo {author} {\bibfnamefont {N.}~\bibnamefont {Di~Bartolomeo}}, \bibinfo {author} {\bibfnamefont {R.}~\bibnamefont {Gatto}}, \bibinfo {author} {\bibfnamefont {F.}~\bibnamefont {Feruglio}},\ and\ \bibinfo {author} {\bibfnamefont {G.}~\bibnamefont {Nardulli}},\ }\bibfield  {title} {\bibinfo {title} {{Phenomenology of heavy meson chiral Lagrangians}},\ }\href {https://doi.org/10.1016/S0370-1573(96)00027-0} {\bibfield  {journal} {\bibinfo  {journal} {Phys. Rept.}\ }\textbf {\bibinfo {volume} {281}},\ \bibinfo {pages} {145} (\bibinfo {year} {1997})},\ \Eprint {https://arxiv.org/abs/hep-ph/9605342} {arXiv:hep-ph/9605342} \BibitemShut {NoStop}%
\bibitem [{\citenamefont {Detmold}\ \emph {et~al.}(2011)\citenamefont {Detmold}, \citenamefont {Lin},\ and\ \citenamefont {Meinel}}]{Detmold:2011rb}%
  \BibitemOpen
  \bibfield  {author} {\bibinfo {author} {\bibfnamefont {W.}~\bibnamefont {Detmold}}, \bibinfo {author} {\bibfnamefont {C.~J.~D.}\ \bibnamefont {Lin}},\ and\ \bibinfo {author} {\bibfnamefont {S.}~\bibnamefont {Meinel}},\ }\bibfield  {title} {\bibinfo {title} {{Axial couplings in heavy hadron chiral perturbation theory at the next-to-leading order}},\ }\href {https://doi.org/10.1103/PhysRevD.84.094502} {\bibfield  {journal} {\bibinfo  {journal} {Phys. Rev. D}\ }\textbf {\bibinfo {volume} {84}},\ \bibinfo {pages} {094502} (\bibinfo {year} {2011})},\ \Eprint {https://arxiv.org/abs/1108.5594} {arXiv:1108.5594 [hep-lat]} \BibitemShut {NoStop}%
\bibitem [{\citenamefont {Workman}\ \emph {et~al.}(2022)\citenamefont {Workman} \emph {et~al.}}]{ParticleDataGroup:2022pth}%
  \BibitemOpen
  \bibfield  {author} {\bibinfo {author} {\bibfnamefont {R.~L.}\ \bibnamefont {Workman}} \emph {et~al.} (\bibinfo {collaboration} {Particle Data Group}),\ }\bibfield  {title} {\bibinfo {title} {{Review of Particle Physics}},\ }\href {https://doi.org/10.1093/ptep/ptac097} {\bibfield  {journal} {\bibinfo  {journal} {PTEP}\ }\textbf {\bibinfo {volume} {2022}},\ \bibinfo {pages} {083C01} (\bibinfo {year} {2022})}\BibitemShut {NoStop}%
\bibitem [{\citenamefont {Epelbaum}\ \emph {et~al.}(2020)\citenamefont {Epelbaum}, \citenamefont {Krebs},\ and\ \citenamefont {Reinert}}]{Epelbaum:2019kcf}%
  \BibitemOpen
  \bibfield  {author} {\bibinfo {author} {\bibfnamefont {E.}~\bibnamefont {Epelbaum}}, \bibinfo {author} {\bibfnamefont {H.}~\bibnamefont {Krebs}},\ and\ \bibinfo {author} {\bibfnamefont {P.}~\bibnamefont {Reinert}},\ }\bibfield  {title} {\bibinfo {title} {{High-precision nuclear forces from chiral EFT: State-of-the-art, challenges and outlook}},\ }\href {https://doi.org/10.3389/fphy.2020.00098} {\bibfield  {journal} {\bibinfo  {journal} {Front. in Phys.}\ }\textbf {\bibinfo {volume} {8}},\ \bibinfo {pages} {98} (\bibinfo {year} {2020})},\ \Eprint {https://arxiv.org/abs/1911.11875} {arXiv:1911.11875 [nucl-th]} \BibitemShut {NoStop}%
\bibitem [{\citenamefont {Baru}\ \emph {et~al.}(2015)\citenamefont {Baru}, \citenamefont {Epelbaum}, \citenamefont {Filin},\ and\ \citenamefont {Gegelia}}]{Baru:2015ira}%
  \BibitemOpen
  \bibfield  {author} {\bibinfo {author} {\bibfnamefont {V.}~\bibnamefont {Baru}}, \bibinfo {author} {\bibfnamefont {E.}~\bibnamefont {Epelbaum}}, \bibinfo {author} {\bibfnamefont {A.~A.}\ \bibnamefont {Filin}},\ and\ \bibinfo {author} {\bibfnamefont {J.}~\bibnamefont {Gegelia}},\ }\bibfield  {title} {\bibinfo {title} {{Low-energy theorems for nucleon-nucleon scattering at unphysical pion masses}},\ }\href {https://doi.org/10.1103/PhysRevC.92.014001} {\bibfield  {journal} {\bibinfo  {journal} {Phys. Rev. C}\ }\textbf {\bibinfo {volume} {92}},\ \bibinfo {pages} {014001} (\bibinfo {year} {2015})},\ \Eprint {https://arxiv.org/abs/1504.07852} {arXiv:1504.07852 [nucl-th]} \BibitemShut {NoStop}%
\bibitem [{\citenamefont {Baru}\ \emph {et~al.}(2016)\citenamefont {Baru}, \citenamefont {Epelbaum},\ and\ \citenamefont {Filin}}]{Baru:2016evv}%
  \BibitemOpen
  \bibfield  {author} {\bibinfo {author} {\bibfnamefont {V.}~\bibnamefont {Baru}}, \bibinfo {author} {\bibfnamefont {E.}~\bibnamefont {Epelbaum}},\ and\ \bibinfo {author} {\bibfnamefont {A.~A.}\ \bibnamefont {Filin}},\ }\bibfield  {title} {\bibinfo {title} {{Low-energy theorems for nucleon-nucleon scattering at $M_\pi=450$ MeV}},\ }\href {https://doi.org/10.1103/PhysRevC.94.014001} {\bibfield  {journal} {\bibinfo  {journal} {Phys. Rev. C}\ }\textbf {\bibinfo {volume} {94}},\ \bibinfo {pages} {014001} (\bibinfo {year} {2016})},\ \Eprint {https://arxiv.org/abs/1604.02551} {arXiv:1604.02551 [nucl-th]} \BibitemShut {NoStop}%
\bibitem [{\citenamefont {Landau}\ \emph {et~al.}(2007)\citenamefont {Landau}, \citenamefont {Paez},\ and\ \citenamefont {Bordeianu}}]{Landautxt}%
  \BibitemOpen
  \bibfield  {author} {\bibinfo {author} {\bibfnamefont {R.}~\bibnamefont {Landau}}, \bibinfo {author} {\bibfnamefont {M.}~\bibnamefont {Paez}},\ and\ \bibinfo {author} {\bibfnamefont {C.}~\bibnamefont {Bordeianu}},\ }\href {https://books.google.de/books?id=RBg-vgAACAAJ} {\emph {\bibinfo {title} {Computational Physics: Problem Solving with Computers}}}\ (\bibinfo  {publisher} {Wiley},\ \bibinfo {year} {2007})\BibitemShut {NoStop}%
\end{thebibliography}%


  \begin{appendix}

\section{Framework}

\subsection{Plane wave expansion method}
Relativistic two-body dynamics can be described in the framework of the Bethe-Salpeter equation. 
To transform it to a numerically more tractable three-dimensional integral equation, we use the method of Blankenbecler and Sugar equation~\cite{Woloshyn:1973mce}.

The Green function reads
\begin{eqnarray}\label{eq:sm_G_sup}
G(\bm{q},E)	&=&i\int\frac{dq^{0}}{2\pi}\frac{1}{(P-q)^{2}-m_{1}^{2}+i\epsilon}\frac{1}{q^{2}-m_{2}^{2}+i\epsilon} \nonumber\\
&=&\frac{1}{4\omega_{1}\omega_{2}} \left(  \frac1{E-\omega_1-\omega_2} - \frac1{E +\omega_1+\omega_2} \right)\nonumber\\
	&=&\frac{1}{2\omega_{1}\omega_{2}}\frac{(\omega_{1}+\omega_{2})}{E^{2}-(\omega_{1}+\omega_{2})^{2}+i\epsilon}\,,
\end{eqnarray}
with $\omega_{i}=\sqrt{m_{i}^{2}+\bm{q}^{2}}$. Here, the total four-momentum of the two-particle system in the center of mass frame is $P^{\mu}=(E,\bm{0})$, while $q$ and $P-q$ refer to the four-momenta of the particles.  The second line in Eq.~\eqref{eq:sm_G_sup} 
corresponds to the propagators written in terms of time-ordered perturbation theory (TOPT), with 
the first term possessing the two-body $DD^*$ cut~\footnote{Neglecting the last term in the second line in Eq.~\eqref{eq:sm_G_sup} corresponds to the Kadyshevsky equation \cite{Kadyshevsky:1967rs} (see also \cite{Baru:2019ndr} for a related discussion). We checked that omitting this term does not have an impact on our results.}.   
The effects neglected in this treatment start from the so-called stretched boxes -- one loop diagrams in TOPT with no two-body cuts involved and therefore contributing at higher orders, see, e.g.,~\cite{Zhang:2021hcl,Hanhart:2000wf} for a related discussion. 
The Lippmann-Schwinger-type three-dimensional equation is
\begin{equation}\label{eq:LSE_supl}
T(\bm{p},\bm{p}'\!,E)=V(\bm{p},\bm{p}')+\!\int\!\!\frac{d^{3}\bm{q}}{(2\pi)^{3}}V(\bm{p},\bm{q})G(\bm{q},E)T(\bm{q},\bm{p}'\!,E).
\end{equation}
The three-dimensional momenta which appear in the equation above are defined in the center of mass frame. 

For two particles in the FV, the momenta become discrete to satisfy the periodic boundary condition. The quantization condition of momenta in the box frame is $\bm{q}_{n}^{b}=\frac{2\pi}{L}\bm{n}$ with $\bm n\in Z^3$. The total three-dimensional momentum in the box frame is $\bm{P}={2\pi \over L}\bm d$ with $\bm d \in Z^3$. To obtain the LSE in the FV, we replace the integration over the loop momentum  with a summation
\begin{equation}
    \int\frac{d^{3}\bm{q}}{(2\pi)^{3}}f(\bm{q})=\int\frac{d^{3}\bm{q}^{b}}{(2\pi)^{3}}{\cal J}f[\bm{q}(\bm{q}^{b})]\to\sum_{\bm{n}\in Z^{3}}\frac{1}{L^{3}}{\cal J}f[\bm{q}(\bm{q}_{n}^{b})],
    \label{eq:sm_temp_01}
\end{equation}
where the Lorentz boost transformation connecting $\bm q$ in the center of mass frame and $\bm q ^b$ in the box frame reads~\cite{Li:2021mob}
\begin{eqnarray}
    \bm{q}	&&=\gamma\left(\bm{q}_{\parallel}^{b}-\frac{\omega_{1}^{b}}{\omega_{1}^{b}+\omega_{2}^{b}}\bm{P}\right)+\bm{q}_{\perp}^{b},\\
\gamma	&&=\frac{\omega_{1}^{b}+\omega_{2}^{b}}{\sqrt{(\omega_{1}^{b}+\omega_{2}^{b})^{2}-\bm{P}^{2}}},
\end{eqnarray}
with $\omega_{1}^{b}=\sqrt{m_{1}^{2}+(\bm{q}^{b})^2}$ and $\omega_{2}^{b}=\sqrt{m_{2}+(\bm{P}-\bm{q}^{b})^{2}}$, 
 $\bm{q}_{\parallel}^{b} =(\bm{q}^{b}\cdot {\bm P}){\bm P}/{\bm P}^2$ and $\bm{q}_{\perp}^{b}=\bm{q}^b-\bm{q}_{\parallel}^{b}$. The Jacobian $\mathcal{J}$ in Eq.~\eqref{eq:sm_temp_01} has the form
\begin{equation}
    {\cal J}=\left(\frac{\omega_{1}\omega_{2}}{\omega_{1}+\omega_{2}}\right)\left(\frac{\omega_{1}^{b}\omega_{2}^{b}}{\omega_{1}^{b}+\omega_{2}^{b}}\right)^{-1}.
\end{equation}
In FV, the LSE turns into the matrix equation
\begin{equation}
    \mathbb{T}=\mathbb{V}+\mathbb{V}.\mathbb{G}.\mathbb{T}\,,
\end{equation}
with
\begin{eqnarray}
\mathbb{G}(E)=\frac{{\cal J}(\bm{q}_{n})}{L^{3}}G(\bm{q}_{n},E)\delta_{\bm{n}',\bm{n}},
\quad
\mathbb{V}	=V(\bm{q}_{n},\bm{q}_{n'})\,.
\end{eqnarray}
The FV energy levels can be obtained by solving the equation 
\be 
\text{det}[\mathbb{G}^{-1}(E)-\mathbb{V}]=0.
\ee
One can introduce the modified Green function and potential matrices
\begin{eqnarray}
&\tilde{\mathbb{G}}^{-1}(E)	=E^{2}\mathbb{I}-\tilde{\mathbb{H}}_{0},\quad
\tilde{\mathbb{V}}	=\frac{1}{L^{3}}\sqrt{\mathbb{N}}.\mathbb{V}.\sqrt{\mathbb{N}},\,
\end{eqnarray}
with
\begin{align}
 &   \tilde{\mathbb{H}}_{0}=[\omega_{1}(\bm{q}_{n})+\omega_{2}(\bm{q}_{n})]^{2}\delta_{\bm{n},\bm{n}'},\\
&\mathbb{N}=\frac{\omega_{1}^{b}(\bm{q}_{n})+\omega_{2}^{b}(\bm{q}_{n})}{2\omega_{1}^{b}(\bm{q}_{n})\omega_{2}^{b}(\bm{q}_{n})}\delta_{\bm{n},\bm{n}'},
\end{align}
 and  identity matrix $\mathbb{I}$, so that the determinant equation to obtain the FV energy levels becomes $\text{det}[\tilde{\mathbb{G}}^{-1}(E)-\tilde{\mathbb{V}}]=0$. The solutions can be found by solving the eigenvector problem
 \begin{equation}
    (\mathbb{\tilde{H}}_{0}+\mathbb{\tilde{V}})\bm{v}=E^{2}\bm{v},~\label{eq:sm_hamilton_eq}
\end{equation}
where $\bm{v}$ is the eigenvector.

The plane wave basis with discrete momenta gives rise to a (reducible) representation of the corresponding point group. We can decompose this representation into irreps using the projection operator technique, which is discussed in detail in Ref.~\cite{Meng:2021uhz}. After reduction, the matrix $\mathbb{H}\equiv \tilde{\mathbb{H}}_0+ \tilde{\mathbb{V}}$ becomes block-diagonal, $\mathbb{H}=\text{diag}(\mathbb{H}^{\Gamma_1},\mathbb{H}^{\Gamma_2},...)$, where $\Gamma_i$ labels different irreps. For a specific irrep, one can obtain the FV energy levels by solving the eigenvalue problem of the sub-matrix, $\mathbb{H}^{\Gamma_i}\bm{v}=E^2\bm{v}$.

In the above derivation, no partial wave expansion is performed. The usage of the plane wave basis is advantageous to include partial wave mixing effect arising from breaking of the rotation symmetry in a cubic box. In the context of moving two-body systems with unequal masses in the FV,  space inversion invariance is also broken,  resulting in a mixture of states with even and odd parities~\cite{Leskovec:2012gb,Doring:2012eu}. For example, for the states residing in the $A_2(1)$ irrep in Fig.~\ref{fig:E_FV} of the main text, the parity is not a good quantum number, since the energy levels in this irrep receive contributions from the $DD^*$ interactions both in $S$ and $P$ waves.

\subsection{Chiral EFT interactions}

For the $DD^*$ system, the OPE constitutes the longest-range interactions (apart from negligibly small electromagnetic interactions). 
We choose the kinematics as illustrated in Fig.~\ref{fig:sm_ope} such that $\bm p$ ($\bm p'$) denotes the momentum of the initial (final) $D^*$ meson
and the corresponding $D$ mesons have momenta of opposite signs.
In this convention, the initial and final relative momenta between $D^*$ an $D$ mesons 
are simply $\bm p$ and $\bm p'$ and have the same sign.  
Such choice is convenient 
when  the partial wave decomposition is performed, since in this case the initial and final states are decomposed in the same way. 
We also introduce the combinations $\bm q= \bm p'-\bm p$ and $\bm k= \bm p'+ \bm p$.
The regularized leading-order (LO) OPE potential is expressed as follows, see Ref.~\cite{Reinert:2017usi} for a related discussion:

\begin{figure}
    \centering
    \includegraphics[width=0.25\textwidth]{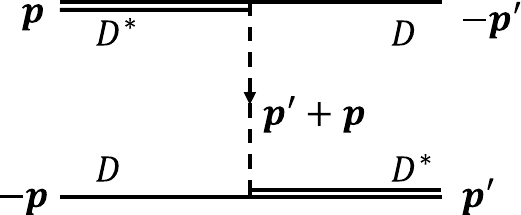}
    \caption{Feynman diagram for one-pion exchange interactions. }
    \label{fig:sm_ope}
\end{figure}
\begin{eqnarray}\label{eq:sm_VOPE_sup}
  && V_{\text{OPE}}^{(0)}(\bm{p}',\bm{p})=C_{\text{iso}}\frac{4M_{D}M_{D^{*}}g^{2}}{4f_{\pi}^{2}}D(\bm{k})e^{-\frac{k^{2}+\mu^2}{\Lambda^{2}}},\\
   && D(\bm{k})=\frac{(\bm{k}\cdot\bm{\epsilon}'^{*})(\bm{k}\cdot\bm{\epsilon})}{\bm{k}^{2}+\mu^2}+C_{\text{sub}} (\bm{\epsilon}'^{*}\cdot\bm{\epsilon}),\\
    &&C_{\text{sub}}=-\frac{\Lambda(\Lambda^{2}-2\mu^{2})+2\sqrt{\pi}\mu^{3}e^{\frac{\mu^{2}}{\Lambda^{2}}}\text{erfc}(\frac{\mu}{\Lambda})}{3\Lambda^{3}}
\end{eqnarray}
where $ \mu^2=m_{\pi}^2-\Delta M^2$,  $\Delta M = M_{D^*}-M_D$  and $\Lambda$ is the cutoff parameter. Furthermore, the isospin coefficient for the isospin singlet $DD^*$ system reads $C_\text{iso}=-3$.
The effective mass $\mu$ sets the typical scale of the OPE interaction in the $D D^*$ system. 
It is worthwhile to stress that for the physical pion mass, one has  $\mu^2<0$. 
 In this case, however, the kinetic energies of $D $ and $D^{*}$, neglected in Eq.~\eqref{eq:sm_VOPE_sup}, 
have to be taken into account. This results in the modification of the static propagator  in the 
OPE potential in the three-body ($DD\pi$) TOPT Green function, which can go on shell. Consequently, in this case one would encounter the three-body ($DD\pi$) cut \cite{Du:2023hlu}.  
For such values of the quark masses,  exponentially suppressed corrections from the OPE and partial wave mixing are expected to become increasingly significant.

Up to ${\cal O}(Q^2)$ (NLO), one can construct six contact operators in general, $\bm{\epsilon}'^{*}\cdot\bm{\epsilon}$, $\bm{q}^2(\bm{\epsilon}'^{*}\cdot\bm{\epsilon})$, $\bm{k}^{2}(\bm{\epsilon}'^{*}\cdot\bm{\epsilon})$, $(\bm{q}\cdot\bm{\epsilon}'^{*})(\bm{q}\cdot\bm{\epsilon})$, $(\bm{k}\cdot\bm{\epsilon}'^{*})(\bm{k}\cdot\bm{\epsilon})$ and $(\bm{\epsilon}'^{*}\times\bm{\epsilon})\cdot(\bm{q}\times\bm{k})$, which are labeled as $\mathcal{O}_{1-6}$ in order. We can recombine them into six operators contributing to specific partial waves,
\begin{eqnarray}
&&{\cal O}_{^{3}S_{1}}^{(0)}={\cal O}_{1},\nonumber\\
&&{\cal O}_{^{3}S_{1}}^{(2)}=\frac{1}{2}{\cal O}_{2}+\frac{1}{2}{\cal O}_{3},\nonumber\\
&&{\cal O}_{^{3}S_{1}-{}^{3}D_{1}}^{(2)}=-{\cal O}_{2}-{\cal O}_{3}+3{\cal O}_{4}+3{\cal O}_{5},\nonumber\\
&&{\cal O}_{^{3}P_{0}}^{(2)}=-\frac{1}{4}({\cal O}_{4}-{\cal O}_{5})+\frac{1}{4}{\cal O}_{6},~\label{eq:sm_all_ctc}\\
&&{\cal O}_{^{3}P_{1}}^{(2)}=-\frac{3}{2}({\cal O}_{2}-{\cal O}_{3})+\frac{3}{2}({\cal O}_{4}-{\cal O}_{5})+\frac{3}{2}{\cal O}_{6},\nonumber\\
&&{\cal O}_{^{3}P_{2}}^{(2)}=-\frac{3}{2}({\cal O}_{2}-{\cal O}_{3})-\frac{1}{2}({\cal O}_{4}-{\cal O}_{5})-\frac{5}{2}{\cal O}_{6}.\nonumber
\end{eqnarray}

 Alternatively, by applying the Fierz transformations, one can express 
the relevant contact operators, corresponding to the diagonal partial-wave transitions
${^{3}S_{1}},{^{3}P_{0}}$ and ${^{3}P_{2}}$, as well as to the off-diagonal transition ${^{3}S_{1}}\to{^{3}D_{1}}$ as (see also Eq.~\eqref{eq:Vct} in the main text)
\bea 
\nonumber
V_{\text{cont}}^{(0)+(2)}[^3S_1]&=&\left(C^{(0)}_{^{3}S_{1}}  +C^{(2)}_{^{3}S_{1}}  (p^{2}+p'^{2})\right )  
P[^3S_1]_i P[^3S_1]_i^\dagger\\ \nonumber\
V_{\text{cont}}^{(2)}[^3\!S_1\!-\!^3\!D_1]&=&  C^{(2)}_{SD}  p'^{2}\;   
P[^3S_1]_i P[^3D_1]_i^\dagger\\
V_{\text{cont}}^{(2)}[^3P_0]&=&C^{(2)}_{^{3}P_{0}}\; 
P[^3P_0] P[^3P_0]^\dagger~\label{eq:sm_Vct_sup},\\
\nonumber
V_{\text{cont}}^{(2)}[^3P_2]&=&C^{(2)}_{^{3}P_{2}}\;  P[^3P_2]_{ij} P[^3P_2]_{ij}^\dagger,
\eea
where $P[^{2S+1}L_J]$ ($P^\dagger[^{2S+1}L_J]$) denotes a projector of the initial (final) $DD^*$ pair onto the partial wave $^{2S+1}L_J$, normalized according to~\cite{Baru:2019xnh}
\be\nonumber
\frac1{2J+1} \int \frac{d\Omega}{4\pi} {\rm Tr}\left( P[^{2S+1}L_J]_{ij\dots} P^\dagger[^{2S+1}L_J]_{ij\dots}\right) =1,
\ee
where the trace is taken with respect to the spin indices $i,j,\dots$\ .  Specifically, 
the relevant projectors read 
\bea\nonumber
P[^3S_1]_i &=& \bm\epsilon_i,\\\nonumber
P[^3D_1]_i &=& -\frac{\sqrt{3}}{2}\bm\epsilon_i \left(\bm n_i \bm n_j -\frac13 \delta_{ij}\right),\\\nonumber
P[^3P_0] &=&  {\bm n \cdot \bm\epsilon  }  , \\\nonumber
P[^3P_2]_{ij} &=&  \frac{\sqrt{3}}{2}\left({\bm n_i\bm\epsilon_j+\bm n_j\bm\epsilon_i- \frac23 ({\bm n \cdot \bm\epsilon  }  ) \delta_{ij}  }  \right), 
\eea
where $\bm n=\bm p/|\bm p|$.

\subsection{Details of the fits to lattice QCD data}
To determine the LECs from the best fit to the FV energy levels obtained from lattice simulations, the following $\chi^2$ function is minimized, 
\begin{align}
\chi^{2}(\{C_{\text{cont.}}\})=	\sum_{i,j}&\delta E_{cm}(i,\{C_{\text{cont.}}\}) \nonumber \\
&\mathcal{C}^{-1}(i,j)\delta E_{cm}(j,\{C_{\text{cont.}}\})\,,
\end{align}
where 
\begin{align}
    \delta E_{cm}(i,\{C_{\text{cont.}}\})=E_{cm}^{\text{LQCD}}(i)-E_{cm}^{\text{EFT}}(i,\{C_{\text{cont.}}\})\,,
\end{align}
is the difference between the FV energy levels from 
lattice QCD simulations ($E_{cm}^{\text{LQCD}}(i)$) and our EFT calculations ($E_{cm}^{\text{EFT}}(i,\{C_{\text{cont.}}\})$).  $\mathcal{C}$ is the covariance matrix of the lattice calculation~\cite{Padmanath:2022cvl}.

\begin{figure}
    \centering
    \includegraphics[width=0.3\textwidth]{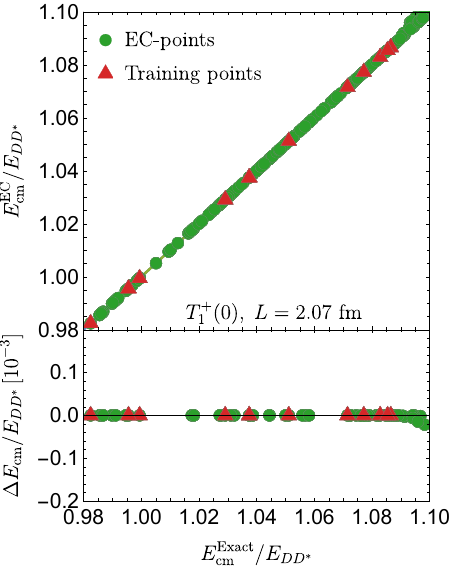}
    \caption{Comparison of the exact results for the FV energy levels $E_{cm}^\text{Exact}$  (green circles labeled as EC-points) and the approximations $E_{cm}^\text{EC}$ obtained using eigenvalue continuation   (red triangles labeled as training points). $\Delta E_{cm}$ is the deviation $\Delta E_{cm}=E_{cm}^\text{Exact}-E_{cm}^\text{EC}$.  
    }
    \label{fig:sm_EC}
\end{figure}

In the fitting and uncertainty quantification procedures, the eigenvalue problem in Eq.~\eqref{eq:sm_hamilton_eq} is repeatedly solved for different values of the LECs. Using the plane wave basis, the dimension of the matrix $\mathbb{H}$  is rather large
\begin{equation}
    N_{\text{Exact}}\sim\left(\Lambda_{\text{UV}}/\frac{2\pi}{L}\right)^{3}\sim \mathcal{O}(1000),
\end{equation}
where $\Lambda_\text{UV}$ is the truncation scale of the discrete momentum modes. In principle, one can choose $\Lambda_\text{UV}$ corresponding to the lattice spacing $a$ as the truncation,  which would result in $\Lambda_\text{UV} \approx 2.3$ GeV. In our calculations, we choose $\Lambda_\text{UV}\approx 4$ GeV to make $\Lambda_\text{UV}$ significantly greater than the regulator $\Lambda$ of the chiral EFT, ensuring that the energy levels $E_{FV}$ are independent of the truncation scale. To accelerate the calculations, we employ a recently developed subspace learning technique known as eigenvector continuation (EC) (see~\cite{Duguet:2023wuh} for a recent review). In this approach, we  randomly choose several sets of LECs, referred to as the training points, and use them to solve the eigenvalue problem exactly during the subspace learning process. Subsequently, the eigenvalue problem for arbitrary LECs is solved in the subspace spanned by the eigenvectors of the training points. It is expected that the solutions in the subspace serve as good approximations of the exact ones, significantly accelerating the calculations due to the drastically reduced dimensionality of the subspace. The dimensionality of the matrix after subspace learning is roughly given by,
\begin{equation}
N_{\text{EC}}\sim\left(p_{\text{max}}/\frac{2\pi}{L}\right)\sim \mathcal{O}(10),
\end{equation}
where $p_{\text{max}}\approx 0.6$ GeV represents the typical momentum of the highest $E_{FV}$ of interest. It is evident that $ p_{\text{max}}<\Lambda\ll\Lambda_{\text{UV}}$.  Notably, $N_{\text{EC}}$ varies linearly with $L$, unlike $N_\text{Exact}$, which increases as $L^3$. Although subspace learning incurs some additional computational cost initially, it is a one-time investment. The reliability of the EC is further facilitated in the EFT framework by using the naturalness assumption to constrain the values of the LECs.  
For a detailed description of this technique we refer to Ref.~\cite{Meng:2023vxy}.

We utilize the finite-volume problem of irreducible representation $T_1^+(0)$ in the box $L=2.07$ fm as an example to showcase the efficiency and accuracy of the EC, as depicted in Fig.~\ref{fig:sm_EC}. Specifically, we select three sets of LECs as training points and consider the first four non-degenerate states to span the subspace. The EC method yields highly accurate results, with discrepancies primarily observed beyond the range covered by the training points and well above the region of interest for $T_1^+(0)$ states, as illustrated in Fig.~ of the main text.

\section{Coupling constants}

To determine the pion decay constant at the unphysical value of the pion mass corresponding to the lattice-QCD simulations of Ref.~\cite{Padmanath:2022cvl} we adopt the values from Ref.~\cite{Du:2023hlu}: $f_\pi=105.3$ MeV for $m_\pi=280$ MeV and $f_{0}=85$ MeV in the chiral limit. These values are obtained through chiral extrapolation using the formula detailed in Refs.~\cite{Baru:2013rta}.
 Note also that no statistically significant  dependence of the pion decay constant on the lattice spacing  was observed in \cite{BMW:2013fzj}.

\begin{figure*}[htp]
    \centering
    \includegraphics[width=0.8\textwidth]{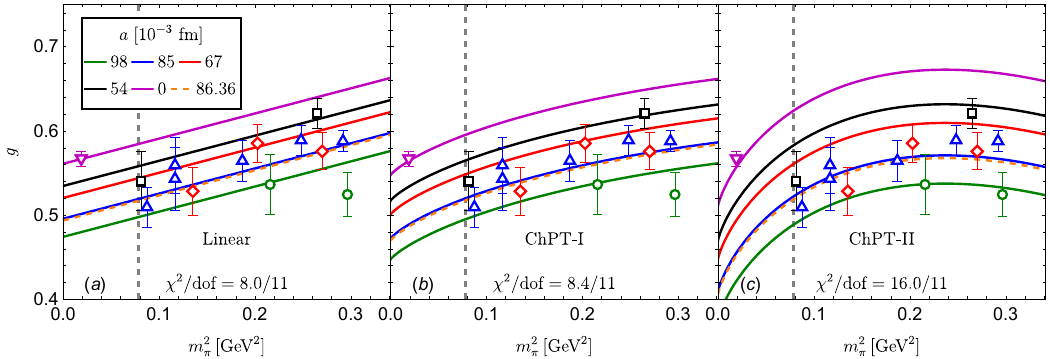}
    \caption{The $D^*D\pi$ coupling constant as a function of $m_\pi^2$ for different values of the lattice spacing, each represented by a different color.  
    The magenta symbols represent the physical value, while the remaining symbols denote coupling constants obtained from lattice simulations at unphysical pion masses $m_\pi$ and lattice spacings $a$~\cite{Becirevic:2012pf}. The left, middle, and right panels correspond to the extrapolation formulas in Eqs.~\eqref{eq:sm_lin_extro}, \eqref{eq:sm_Ch1_extro}, and \eqref{eq:sm_Ch2_extro}, respectively. The value of the $D^*D\pi$ coupling constant employed in our analysis, which corresponds to the lattice spacing of $a=0.08636$~fm and the pion mass of $m_\pi=280$~MeV used in \cite{Padmanath:2022cvl}, can be read off from the intersection of the orange dashed and gray dashed lines. 
     }
    \label{fig:sm_chiral_extra.}
\end{figure*}

For the $D^*D\pi$ coupling constant, we follow a slightly different procedure as compared to that in Ref.~\cite{Du:2023hlu}. Specifically, to extract this coupling, we perform two-dimensional fits of lattice data~\cite{Becirevic:2012pf} 
by simultaneously varying $m_\pi$ and  
the lattice spacing, since the finite-volume energy levels discussed in the main text are obtained not in the continuum, but at $a=0.08636$~fm. We adopt three different extrapolation formulas given in Ref.~\cite{Becirevic:2012pf}, namely one linear extrapolation and two extrapolations based on chiral perturbation theory (ChPT)~\cite{Casalbuoni:1996pg,Detmold:2011rb}:
\begin{itemize}
 \item Linear extrapolation:
\begin{equation}
    g(a,m_{\pi})=g_{0}(1+\alpha m_{\pi}^{2}+\beta a^{2})\,,~\label{eq:sm_lin_extro}
\end{equation}
\item ChPT-I:
\begin{equation}
g(a,m_{\pi})=g_{0}\left(1-\frac{2g_{0}^{2}}{(4\pi f_{0})^{2}}m_{\pi}^{2}\ln m_{\pi}^{2}+\alpha m_{\pi}^{2}+\beta a^{2}\right)\,,~\label{eq:sm_Ch1_extro}
\end{equation}
\item ChPT-II:
\begin{equation}
    g(a,m_{\pi})=g_{0}\left(1-\frac{1+2g_{0}^{2}}{(4\pi f_{0})^{2}}m_{\pi}^{2}\ln m_{\pi}^{2}+\alpha m_{\pi}^{2}+\beta a^{2}\right)\,.~\label{eq:sm_Ch2_extro}
\end{equation}
   \end{itemize}
 In the two ChPT formulas, the renormalization-scale dependence of the logarithmic term and the counter-term $\propto \alpha$ cancel each other and are, therefore, not shown.  
 For each extrapolation approach, there are three parameters to be determined, namely the coupling constant $g_0$ in the chiral and continuum limits, the coefficient $\alpha$ of the $m_\pi^2$-term and the coefficient $\beta$ controlling the continuum extrapolation. We utilize the lattice data for $g$ at different pion masses and lattice spacings in Ref.~\cite{Becirevic:2012pf} as input. Additionally, the physical value $g^{ph}=g(0,m^{ph}_{\pi})=0.567\pm0.009$, which is determined from the experimental $D^*\to D\pi$ decay width~\cite{ParticleDataGroup:2022pth}, is also used to constrain the three parameters.  
  In Fig.~\ref{fig:sm_chiral_extra.},  the input data are compared with the best-fit results shown as a function of the pion mass at different lattice spacings.
 In Table~\ref{tab:sm_coupling}, the parameters corresponding to the best fits, their uncertainties and the extracted coupling constants at $m_\pi=0.280$~GeV and $a=0.08636$~fm are listed for three different extrapolation methods from Eqs.~\eqref{eq:sm_lin_extro}-\eqref{eq:sm_Ch2_extro}.
 The obtained results for the $D^*D\pi$ coupling constant $g\equiv g(0.08636,0.280)$ from
  different extrapolations are in excellent agreement with each other. 
 In the main text, we adopt the value of $0.517\pm0.015$, which is approximately 20\% smaller than the value used in Ref.~\cite{Du:2023hlu}.

\begin{table}[htp]
    \centering
\begin{tabular*}{\hsize}{@{}@{\extracolsep{\fill}}lcccc@{}}
\hline \hline
 & $g_{0}$ & $\alpha\,[\text{GeV}^{-2}]$ & $\beta\,[\text{fm}^{-2}]$ & $g$\tabularnewline
\hline 
Linear & $0.561(9)$ & $0.53(13)$ & $-16.1(44)$ & $0.517(15)$\tabularnewline
ChPT-I & $0.547(8)$ & $0.24(14)$ & $-19.1(45)$ & $0.517(15)$\tabularnewline
ChPT-II & $0.511(8)$ & $-0.59(15)$ & $-27.6(48)$ & $0.519(15)$\tabularnewline
\hline \hline
\end{tabular*}
    \caption{ The parameters of the considered extrapolations from Eqs.~\eqref{eq:sm_lin_extro}-\eqref{eq:sm_Ch2_extro} obtained from best fits to input data, as described in text, and the extracted values of the coupling constant $g$ at $m_\pi=0.280$~GeV and $a=0.08636$~fm.}
    \label{tab:sm_coupling}
\end{table}

\section{Systematic uncertainties}

\begin{figure}[htp]
    \centering
    \includegraphics[width=0.4\textwidth]{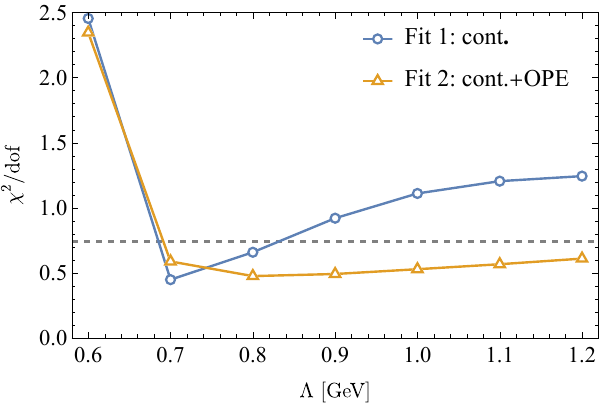}
    \caption{
    Cutoff dependence of the reduced $\chi^2$ for different fits.
     In our Fits 1 and 2, there are three parameters and six degrees of freedom (dof). The gray dashed line represents the reduced $\chi^2$ of the four-parameter fit using LQCs and ERE in Ref.~\cite{Padmanath:2022cvl}. }
    \label{fig:sm_chisqure}
\end{figure}

 In this section, we provide a qualitative discussion of the systematic uncertainties in the results arising from variations in the cutoff in the regularized potentials and the inclusion of additional contact terms at ${\cal O}(Q^2)$. 

The variation of reduced $\chi^2$ values for Fits 1 and 2 is illustrated in Fig.~\ref{fig:sm_chisqure} as the cutoff $\Lambda$ is gradually increased from $0.6$ to $1.2$ GeV with a step size of $0.1$ GeV. Note that the reduced $\chi^2$ for $\Lambda=0.6$ GeV is significantly larger than for other values. However, this observation is not surprising, given that the highest energy level in the fitting data set, corresponding to the $A_1^-(0)$ irrep for $L=2.07$ fm (see Fig.~\ref{fig:E_FV} of the main text), corresponds to a momentum of approximately $0.56$ GeV, which is only slightly smaller than the chosen cutoff.
When $\Lambda$ is varied from $0.7$ to $1.2$ GeV, Fit 2 demonstrates essentially cutoff-independent behavior, while the reduced $\chi^2$ for Fit 1 gradually increases. The reason behind this lies in the fact that in Fit 1, where only one LEC is employed to fit the data for the $^3P_0$ partial wave, the effective range  is primarily provided by the regulators, making the $\chi^2$ sensitive to the choice of cutoff.
In contrast,  Fit 2, with the same number of parameters, exhibits cutoff independence because in this case the range corrections in the $^3P_0$ channel are naturally driven by the OPE interactions.
Another noteworthy observation from Fig.~\ref{fig:sm_chisqure}  is that the reduced $\chi^2$ in our  pionful Fit 2, involving  three parameters, is smaller than that of the four-parameter fit using L\"uscher quantization conditions and the ERE in Ref.~\cite{Padmanath:2022cvl} with the same input for the energy levels. 
{ Indeed, once the OPE is incorporated into the fits, besides its significant influence in the $^3S_1$ channel, the parameter-free long-range physics largely dictates the energy dependence of the $^3P_0$ phase shift. Consequently,   
the 
subleading contact term in the $^3P_0$  channel, needed in Ref.~\cite{Padmanath:2022cvl},  becomes obsolete. The suppression of this contact term, which in chiral EFT enters at $O(Q^4)$,  is fully in line with the hierarchy of the operators   and provides further  support for the whole framework. }

\begin{figure*}[tp]
\begin{center}
\includegraphics[width=0.9\textwidth]{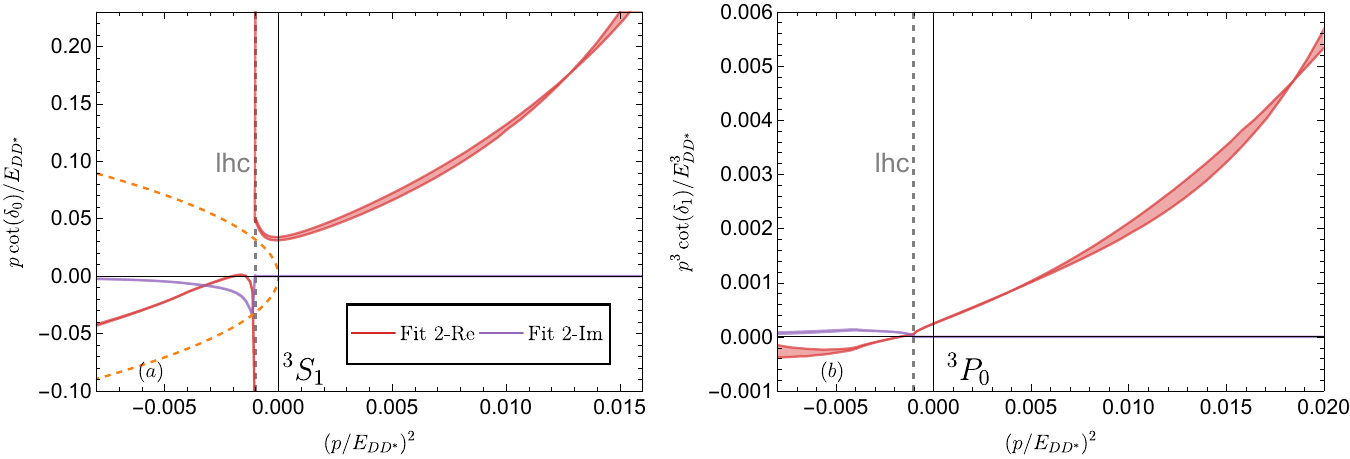}
\caption{ 
Residual cutoff dependence of the  $^3S_1$ (left panel) and $^3P_0$ (right panel) phase shifts corresponding to the cutoff variation $\Lambda=0.7-1.2$ GeV in the pionful Fit 2.
}\label{fig:sm_cut_depend} 
\end{center}
\end{figure*}

To further investigate the effect of varying the cutoff on observables in our  pionful Fit 2, in Fig.~\ref{fig:sm_cut_depend} we present the phase shifts obtained from the best fit while varying cutoff  $\Lambda$ from 0.7 to 1.2 GeV. It can be observed that the systematic uncertainties arising from the cutoff dependence  are very small.

\begin{figure*}[tp]
\begin{center}
\includegraphics[width=0.9\textwidth]{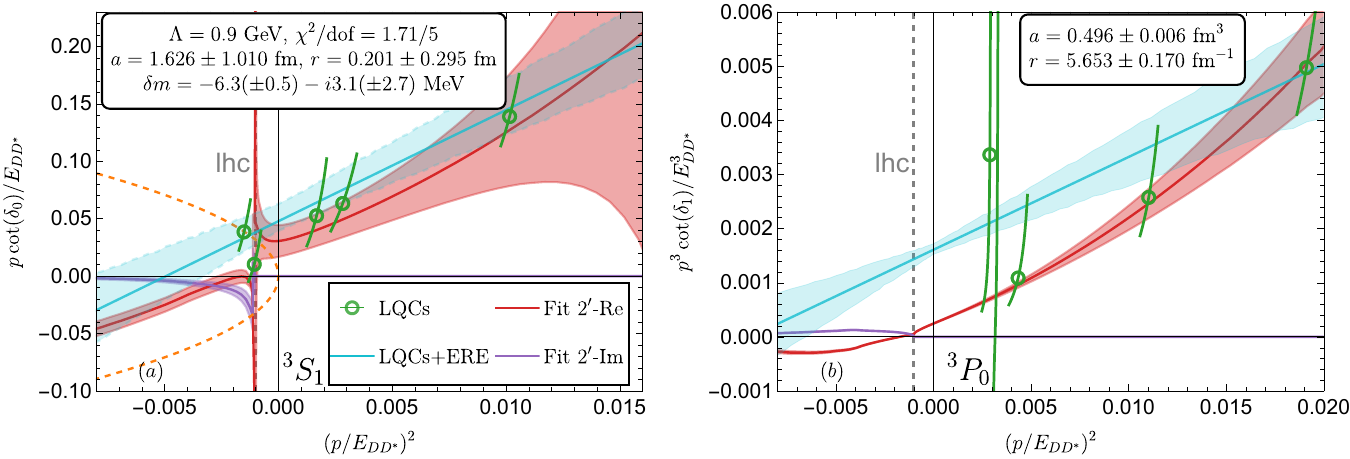}
\caption{\label{fig:sm_SD}  
Phase shifts in the $^3S_1$ (left panel) and $^3P_0$ (right panel) partial waves with an additional ${}^3\!S_1\!-\!{}^3\! D_1$ contact term   
$V_{\text{cont}}^{(2)}[^3\!S_1\!-\!^3\!D_1]$
in Eq.~\eqref{eq:sm_Vct_sup}. The notations are the same as those in Fig.~\ref{fig:phase_shifts} of the main text. The legend shows the ERE parameters and the pole position of the $T_{cc}$. Calculations are performed for the cutoff $\Lambda=0.9$ GeV. }
\end{center}
\end{figure*}

 In our main results, we consider three contact terms:  $V_{\text{cont}}^{(0)+(2)}[^3S_1]$ and $V_{\text{cont}}^{(2)}[^3P_0]$ in Eq.~\eqref{eq:sm_Vct_sup} which 
correspond to the structures   $\mathcal{O}^{(0)}_{^{3}S_{1}}$, $\mathcal{O}^{(2)}_{^{3}S_{1}}$, and $\mathcal{O}^{(2)}_{^{3}P_{0}}$ in Eq.~\eqref{eq:sm_all_ctc}. 
This is justified, since the ${^{3}S_{1}}$ and  ${^{3}P_{0}}$ partial waves are expected to give a dominant contribution to the  energy levels in the irreps $T_1^+(0)$ and  $A_1^-(0)$, respectively.  
Nevertheless,   the partial waves $^3D_1$ and $^3P_2$ also contribute to the FV energy levels used in fits, as indicated in Fig.~\ref{fig:E_FV} of the letter.
To investigate their impact, we conduct two four-parameter fits. In Fit $2'$,  we start from our original pionful formulation corresponding to Fit 2 and supplement 
it with the contact term  $V_{\text{cont}}^{(2)}[^3\!S_1\!-\!^3\!D_1]$ in Eq.~\eqref{eq:sm_Vct_sup}.   
To obtain Fit $2''$, the original Fit 2  is  supplemented with the  contact term  $V_{\text{cont}}^{(2)}[^3P_2]$.   
Then, the four-parameter fits are performed to obtain the best fits to the energy levels as before. 

The results of Fit $2'$, which incorporates the $V_{\text{cont}}^{(2)}[^3\!S_1\!-\!^3\!D_1]$  
contact term, are presented in Fig.~\ref{fig:sm_SD}.
Consequently,  the reduced $\chi^2$   decreases slightly from $2.95/6\approx 0.49$ in Fit 2 to $1.71/5\approx 0.34$ in Fit $2'$.  Meanwhile, the scattering length and effective range in the $^3P_0$ partial 
wave are almost unaffected by the additional contact term, cf. the ERE parameters in Table~\ref{tab:ere_pole_num} of the main text with those shown in the legend in Fig.~\ref{fig:sm_SD}. 
 Also, the change in the pole position of the $T_{cc}$ state is minor and falls within the quoted errors. 
 The central values of the scattering length and effective range of the $^3S_1$ partial wave  experience moderate changes. 
 This is expected, since the ERE has a very limited applicability range, given the presence of the nearby left-hand cut and the pole in $p \cot\delta$. Consequently, the ERE parameters may undergo adjustments when additional interactions are introduced in fits. However, we stress that the modifications in the ERE parameters are consistent with the Fit 2 results within errors. 
 
\begin{figure*}[tp]
\begin{center}
\includegraphics[width=0.9\textwidth]{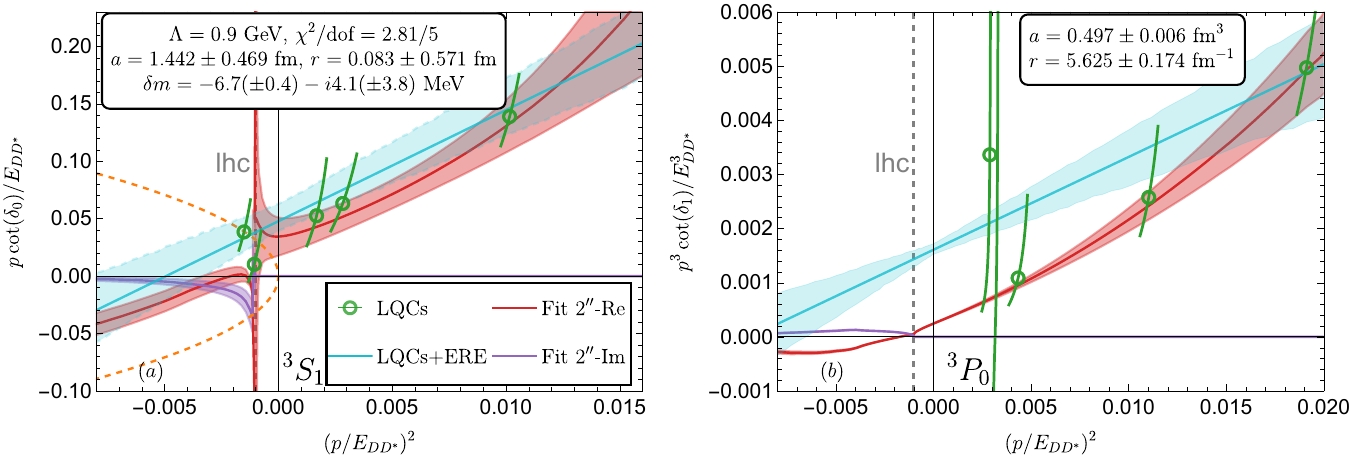}
\caption{\label{fig:sm_3P2}  
Phase shifts in the $^3S_1$ (left panel) and $^3P_0$ (right panel) partial waves with an additional  $^3P_2$ contact term  
$V_{\text{cont}}^{(2)}[^3\!P_2]$
in Eq.~\eqref{eq:sm_Vct_sup}.
 The notations are the same as those in Fig.~\ref{fig:phase_shifts} of the main text. The legend shows the ERE parameters and the pole position of the $T_{cc}$. Calculations are performed for the cutoff $\Lambda=0.9$ GeV.}
\end{center}
\end{figure*}

 The results of Fit $2''$, which incorporates  $V_{\text{cont}}^{(2)}[^3P_2]$,   
 are presented in Fig.~\ref{fig:sm_3P2}. In comparison with the results of Fit 2, the change in $\chi^2$ is very minor, resulting in a slightly larger reduced $\chi^2$. The scattering lengths, effective ranges, and the $T_{cc}$ pole position remain almost unaffected. Therefore, it can be concluded that the effect of the $V_{\text{cont}}^{(2)}[^3P_2]$  term can be neglected given the uncertainties of the present input for the energy levels.
 
 If the accuracy of the energy levels is  improved, one could employ the Bayesian methods to quantify systematic uncertainties for the observables in a more rigorous way.

\section{The left-hand cut problem}

In this section, we discuss the  
 details of the problem  in lattice QCD, related with the emergence of the left-hand cut, which we refer to as the left-hand cut problem. In particular, we  discuss why the left-hand cut in our framework does not pose any technical problems.

{ The  left-hand cut problem occurs when the finite volume energies calculated in lattice QCD are connected with the infinite volume amplitudes using the Lüscher formula~\cite{Green:2021qol,Dawid:2023jrj,Raposo:2023oru}}. 
It is the on-shell infinite volume amplitude $T(E)$,   see  Eq.~\eqref{eq:LuscherQC} in the main text,  which encounters the left-hand cut from the long-range interaction,  restricting the applicability of 
Lüscher's method. Indeed, since the amplitude $T(E)$ 
is complex below the lhc (for details, see below), while the function $F^{-1}(L,\bm{P},E)$  is not, Lüscher's quantization conditions can not be used at least below the lhc.  

{ At this place a comment is in order. 
 The process of extracting physical properties of two-particle scattering from lattice QCD energy levels involves two key steps, as shown in Fig.~\ref{fig:diff}: 
 \begin{itemize}
     \item[] (i) Extracting infinite volume amplitudes (and thus the phase shifts) from the finite volume energy levels. To carry out this step, the lattice calculations so far have used the L\"uscher formula.
     \item[] (ii) Appropriate parameterization of the extracted infinite volume phase shifts to get access to the low-energy scattering properties, such as the effective range parameters,   pole positions and so on.
 \end{itemize}
 Ref.~\cite{Du:2023hlu} demonstrated that the left-hand cut, located in the vicinity of  the threshold, significantly constrains the validity range of the conventional parameterization of the infinite-volume amplitude using the effective range expansion, as employed in step (ii).
However, Ref.~\cite{Du:2023hlu} does not call into question step (i), simply relying on the phase shifts extracted from finite volume energy levels using the L\"uscher method. 
The solution to the lhc problem is proposed in this work (see also \cite{Raposo:2023oru}) and explicitly applied to real lattice data.
\begin{figure}[h]
    \centering
    \includegraphics[width=0.45\textwidth]{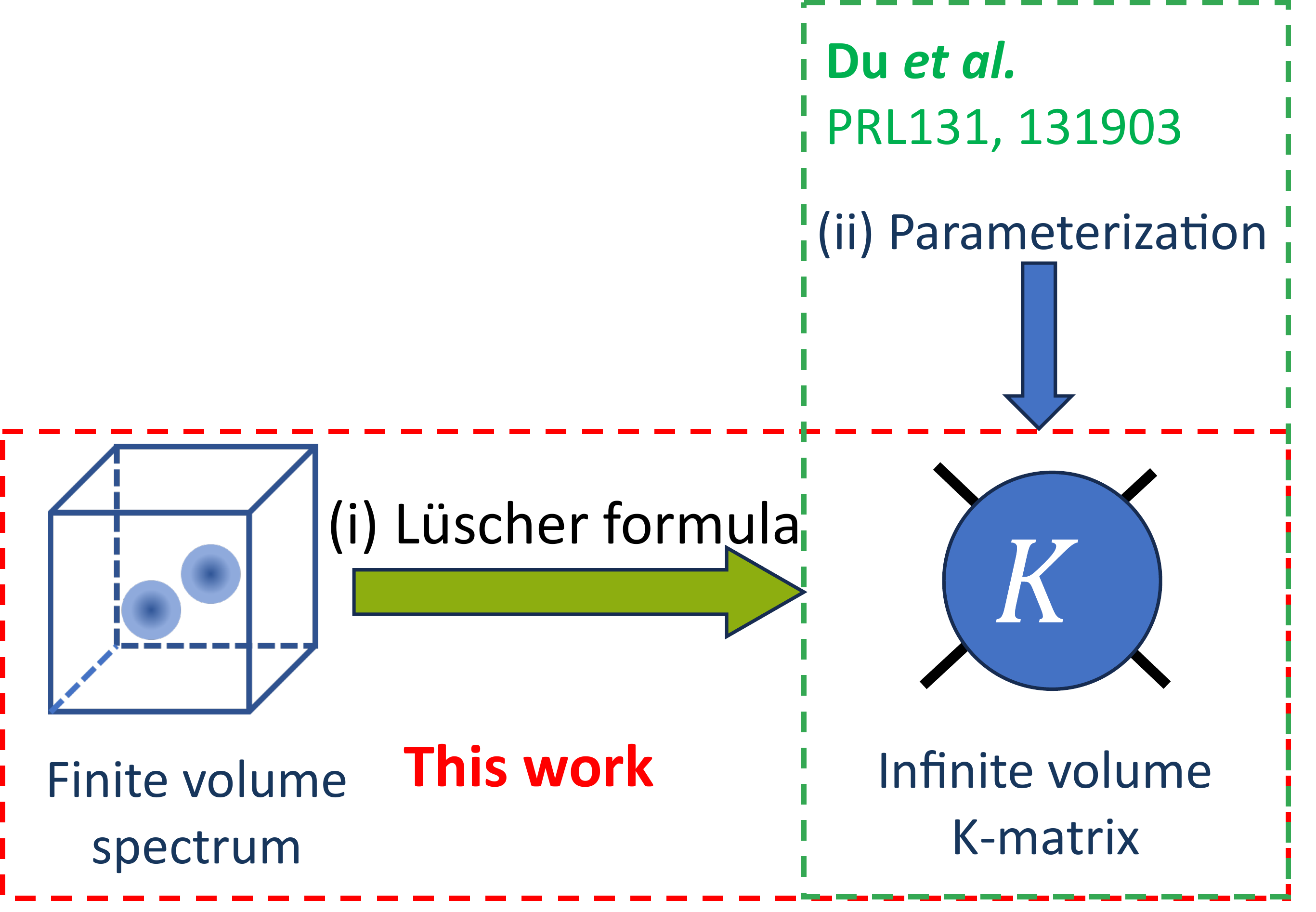}
    \caption{{ Two key steps for 
    extracting physical properties of two-particle scattering
from lattice QCD energy levels. 
   The issues addressed in this work and  Ref.~\cite{Du:2023hlu} are delineated by red and green dashed lines, respectively.
    } 
    }
    \label{fig:diff}
\end{figure}}

In order to illustrate the left-hand cut problem, we write the long-range potential in the form
\begin{align}
V(\bm{p},\bm{p}') & =\frac{f(\bm{p},\bm{p'})}{(\bm{p}+\bm{p}')^{2}+m^{2}}=\frac{f(\bm{p},\bm{p'})}{(p^{2}+p'^{2}+2pp'z)+m^{2}}\,,\label{eq:sm_ope}
\end{align}
where $m$ represents the mass of the exchanged  particle, $\bm{p}$ and $\bm{p'}$ are the off shell momenta and $z=\hat{p}\cdot\hat{p}'$. We also introduce the on shell momentum $p_{\rm on}$, which in the non-relativistic 
kinematics is related to the energy as $p_{\rm on}^2= 2\mu E$ with $\mu$ being the reduced mass of the two-body system. (Please note also that in Eq.~\eqref{eq:ERE} of the main text, the on shell momentum is defined as $p$.) Since the  function  ${f(\bm{p},\bm{p'})}$ in the numerator is smooth and analytic,   to tackle the left-hand cut problem, it can be safely ignored. Therefore, without losing generality, in what follows, we set ${f(\bm{p},\bm{p'})} =1$. 

\begin{figure}
    \centering
    \includegraphics[width=0.35\textwidth]{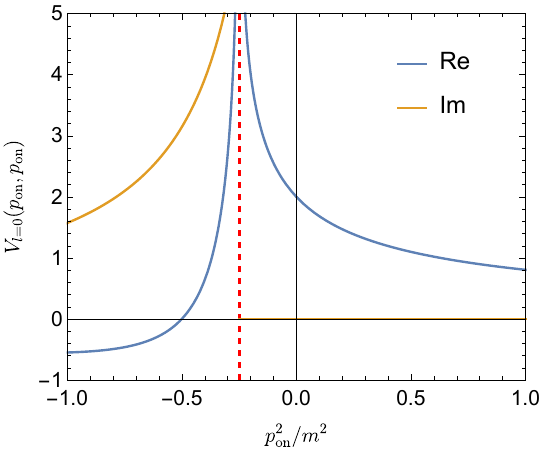}
    \caption{The S-wave on-shell potential, as defined in Eq.~\eqref{eq:Vswave_supl} with $m=m_{\pi}$. The red vertical dashed line denotes the position of the branch point of the left-hand cut at $p_{\rm on}^{2}/m^2=-\frac{1}{4}$. Below this point, the potential, so as the corresponding $T$-matrix, become complex.  The real and imaginary parts of the potential are shown by blue and orange lines, respectively.  
    } 
    \label{fig:sm_Vlhc}
\end{figure}

When $E>0$, there is no singularity arising from the potential in Eq.~\eqref{eq:sm_ope}. Indeed, 
both the on-shell potential $V(p_{\rm on}\bm{n},p_{\rm on}\bm{n'})\equiv V(p_{\rm on},p_{\rm on},z)$  
and the half-off-shell potential $V(p_{\rm on},q,z)$, entering the Lippmann-Schwinger-type equations \eqref{eq:LSE_supl}, are analytic in the domain $E>0$ and $z\in[-1,1]$. 
However, when $E<0$, which corresponds to imaginary $p_{\rm on}$, a singularity arises in the on-shell potential $V(p_{\rm on},p_{\rm on},z)$, since   the denominator of the potential can be equal to zero \be\label{eq:Denom_supl}
2p_{\rm on}^{2}(1+z)+m^{2}=0 \Rightarrow z=-\frac{m^{2}}{2p_{\rm on}^{2}}-1\, ,\
\ee
provided that
\be\label{eq:pon}
 p_{\rm on}^{2}\le-\frac{m^{2}}{4}  \quad \mbox{when}  -1\le z\le1.  
\ee
{ Thus, the on-shell potential  in the plane wave basis has a pole in line with Eqs.~\eqref{eq:Denom_supl}-\eqref{eq:pon}. }
If the partial wave  decomposition is performed, one finds, e.g., for the S-wave potential 
\begin{align}\label{eq:Vswave_supl}
V_{l=0}(p,p') & =\int_{-1}^{1}dz\frac{1}{p^{2}+p'^{2}+2pp'z+m^{2}} \nonumber\\
 & =\frac{1}{2pp'}\log\left(\frac{(p+p')^{2}+m^{2}}{(p-p')^{2}+m^{2}}\right).
\end{align}
{ Instead of a pole in $V(p_{\rm on},p_{\rm on}, z)$, the  on-shell potential $V_{l=0}(p_{\rm on},p_{\rm on})$ in the partial wave basis} 
develops the left-hand cut for $p_{\rm on}^{2}\le-\frac{m^{2}}{4}$,
see Fig.~\ref{fig:sm_Vlhc} for illustration. 
The same is also true  for the corresponding on shell $T$-matrix $T_{l=0}(p_{\rm on},p_{\rm on})$.
Consequently, because of the lhc, the original Lüscher formula cannot be straightforwardly extended to accommodate long-range interactions.

However, it appears crucial to emphasize that, contrary to the on-shell potential, the half off-shell potential $V(p_{\rm on},q,z)$ in Eq.~\eqref{eq:sm_ope}, so as the partial-wave projected potential 
$V_{l=0}(p_{\rm on},q)$  in Eq.~\eqref{eq:Vswave_supl},   with real positive $q$, as used in the LSE,  
{ are free of the pole and the lhc, respectively,  when $E<0$. The same is of course also true for 
the fully off shell potential. }
This is the reason why no problems occur when solving the eigenvalue problem for $E<0$ within our method.

Our procedure can be formulated as a two-step process. In  step 1, we utilize our effective potential,   consisting of the OPE and a series of   contact interactions, to calculate the FV energy levels  both below and above the lhc from the condition that  the
determinant in the LSE vanishes.  
{ Technically, in our case, this amounts to solving the standard eigenvalue problem in the Hamiltonian approach, see Eq.~\eqref{eq:sm_hamilton_eq},
which is completely straightforward,  
 since  the off-shell $t$-channel potential entering Eq.~\eqref{eq:sm_hamilton_eq} only needs to be evaluated for real positive momenta. In this regime, the potential is completely analytic and free of  $t$-channel poles in the plane wave basis (or the $t$-channel lhc in the partial-wave basis).
}
Then, we adjust the low-energy constants   to achieve the best fit to the FV energy levels.  
In  step 2, we use the fully determined  effective potential, with all  low-energy 
constants  fixed  at  step 1,  to calculate the   infinite volume amplitude $T(E)$ from the LSE. 
How to deal with the last step technically is well known from phenomenology of two-body scattering at low-energies, e.g., from $NN$ scattering, where the effects from the lhc are included at the level of multi-pion exchanges, see, e.g., \cite{Epelbaum:2019kcf} for a recent review and \cite{Baru:2015ira,Baru:2016evv} for a related discussion.  The technique of solving the LSE in the infinite volume can be found, e.g. in the textbook~\cite{Landautxt}.

  \end{appendix}


\end{document}